\documentclass[prb,aps,twocolumn]{revtex4}
\usepackage{epsfig,epsf,psfrag}
\usepackage{graphicx}
\usepackage{epic,eepic}
\usepackage{color,pstcol}
\begin{document}
\title{Partially resummed perturbation theory for multiple Andreev reflections in a short
  three-terminal Josephson junction}

\date{\today}
\author{R\'egis M\'elin}

\affiliation{CNRS, Institut NEEL, BP 166,
  F-38042 Grenoble Cedex 9, France}

\affiliation{Universit\'e Grenoble-Alpes, Institut NEEL, BP 166,
  F-38042 Grenoble Cedex 9, France}

\author{Denis Feinberg}

\affiliation{CNRS, Institut NEEL, BP 166,
  F-38042 Grenoble Cedex 9, France}

\affiliation{Universit\'e Grenoble-Alpes, Institut NEEL, BP 166,
  F-38042 Grenoble Cedex 9, France}

\author{Beno\^{\i}t Dou\c{c}ot}

\affiliation{Laboratoire de Physique Th\'eorique et des Hautes
  Energies, CNRS UMR 7589, Universit\'e Pierre et Marie Curie,
  Sorbonne Universit\'es, 4 Place Jussieu, 75252 Paris Cedex 05}

\begin{abstract}

In a transparent three-terminal Josephson junction, modeling
nonequilibrium transport is numerically challenging, owing to the
interplay between multiple Andreev reflection (MAR) thresholds and
multipair resonances in the pair current. An approximate method,
coined as ``partially resummed perturbation theory in the number of
nonlocal Green's functions'', is presented that can be operational on
a standard computer and demonstrates compatibility with results
existing in the literature. In a linear structure made of two
neighboring interfaces (with intermediate transparency) connected by a
central superconductor, tunneling through each of the interfaces
separately is taken into account to all orders. On the contrary,
nonlocal processes connecting the two interfaces are accounted for at
the lowest relevant order. This yields logarithmically divergent
contributions at the gap edges, which are sufficient as a
semi-quantitative description. The method is able to describe the
current in the full two-dimensional voltage range, including
commensurate as well as incommensurate values. The results found for
the multipair (for instance quartet) current-phase characteristics as
well as the MAR thresholds are compatible with previous results.  At
intermediate transparency, the multipair critical current is much
larger than the background MAR current, which supports an experimental
observation of the quartet and multipair resonances. The paper
provides a proof of principle for addressing in the future the
interplay between quasiparticles and multipairs in four-terminal
structures.
\end{abstract}
\maketitle

\begin{figure}[htb]
\includegraphics[width=1\columnwidth]{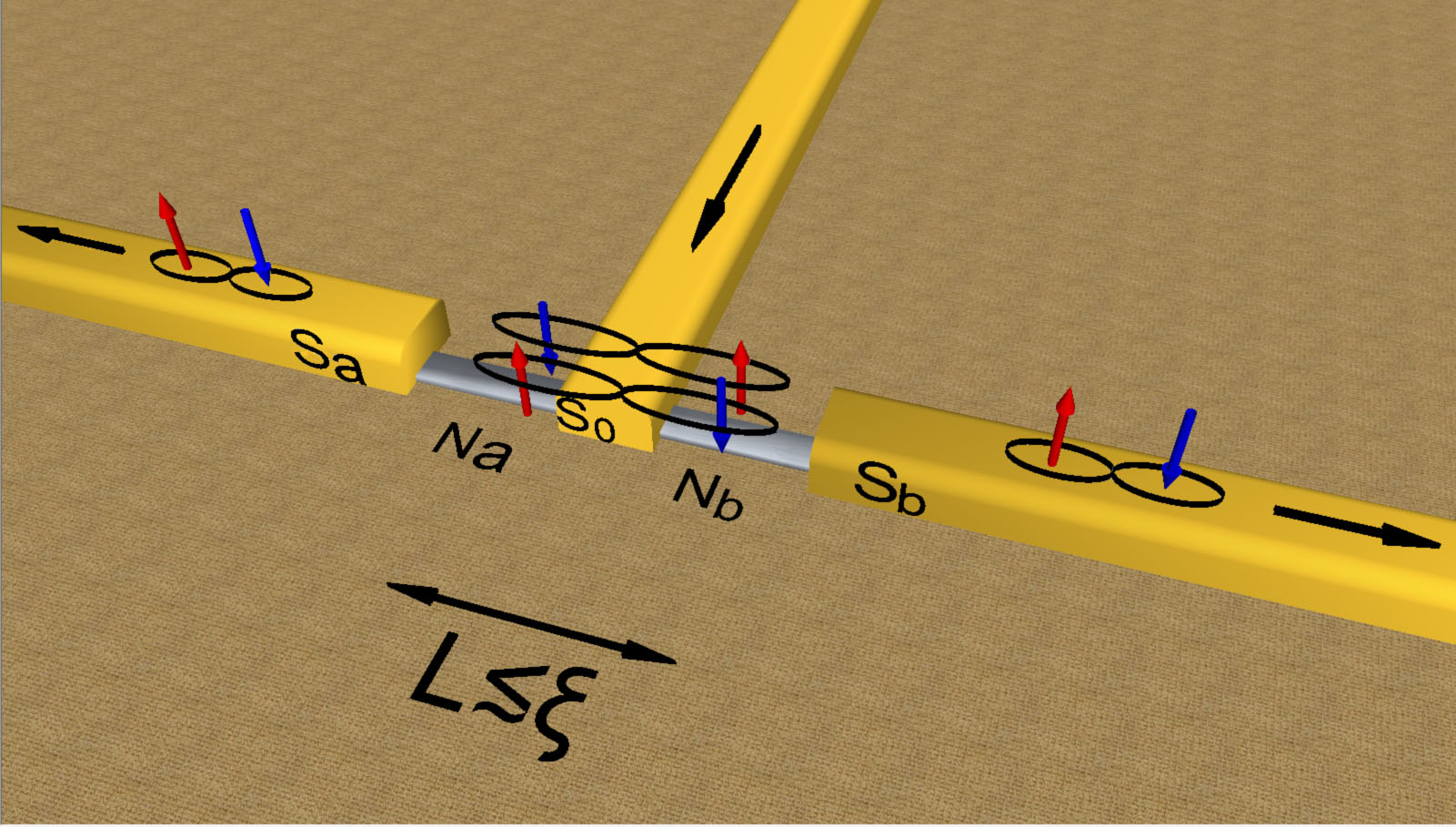}

\caption{Artist view of a three-terminal Josephson
  junction, featuring the quartet resonance. $N_a$ and $N_b$ are two short
  metallic regions, small compared to the BCS coherence length~$\xi$.
\label{fig:artist-view}
}
\end{figure}
  
\section{Introduction}

Multiterminal Josephson junctions have recently focused interest, as a
generalization of standard two-terminal junctions. In particular,
all-superconducting structures with three BCS superconductors $S_a$,
$S_b$ and $S_0$ subject to voltages $V_a$, $V_b$ and $V_0$ (set to
$V_0=0$ by convention) respectively (see Fig.~\ref{fig:artist-view})
have been considered and revealed several types of resonances and
threshold resonances if the DC   currents are calculated as a function
of voltages. The independent voltages $V_a$ and $V_b$ can be
commensurate or incommensurate. For the realistic case of extended
interfaces, the existence of two independent Josephson frequencies
explains the absence of numerically exact results to this
nonequilibrium transport problem. Yet, a number of new physical
effects were obtained theoretically in different limiting cases
(coherent or incoherent limit, ballistic or diffusive) and in different
set-ups (metallic junctions or quantum dots, two or three interfaces).

The starting point was the discovery of what was called ``self-induced
Shapiro steps'' by Cuevas and Pothier\cite{Cuevas-Pothier}. This work
was based on Usadel equations for a diffusive normal metal region
connected to three superconductors, one of them involving a tunnel
contact. DC resonances in the current were obtained at commensurate
voltages. The analogy with Shapiro steps implies a zero-frequency
mode-locking of the Josephson AC oscillations induced at the two
transparent contacts. This reminds of the phenomenon studied in the
$'80s$ in well-separated junctions coupled by a non-mesoscopic
environment\cite{Jillie}. Later on, those resonances were rediscovered
independently by Freyn {\it et al.}\cite{Freyn}. A physical
interpretation was uncovered in this work \cite{Freyn} in terms of
correlations among Cooper pairs in an intermediate virtual state
located in the junction, extending within the coherence length in the
superconducting leads. It was shown that these virtual states of
Cooper pairs lead to entanglement if quantum dots are inserted into
the three-terminal Josephson junction. The prototypical case involves
opposite bias voltages $V_a=-V_b$ (with $V_0=0$) on terminals $S_a$
and $S_b$, which produces correlations among four fermions due to the
exchange between two Cooper pairs (the so-called quartets, see
Fig.~\ref{fig:artist-view}). More generally, correlations among
several Cooper pairs, coined as ``multipair correlations'' appear for
commensurate $V_a$ and $V_b$ ({\it e.g.} if $V_a/V_b$ is a rational
fraction). However, correlations among large numbers of Cooper pair
are damped exponentially if the interfaces are not very transparent,
because an additional electron-hole conversion amplitude crossing
twice the interfaces is required to incorporate one more pair into a
correlated Cooper pair cluster. Multipair correlations and related
phenomena were explored recently by Jonckheere {\it et
  al.}\cite{Jonckheere} for a double quantum dot connected to three
superconductors. The lowest order quartet mode was understood in terms
of splitting two Cooper pairs and recombining them in a way that
involves an exchange between fermions. This mechanism leads to a minus
sign in the phase dependence of the quartet current and is not
captured by the heuristic "synchronization" mechanism assumed by the
Shapiro step analogy (the phase dependence was not studied in
Ref.\onlinecite{Cuevas-Pothier}). Interestingly, another original
process, coined as ``phase-MAR'' was discovered \cite{Jonckheere},
corresponding to an interference term for multiple Andreev reflections
(MARs), with transport of quasiparticles from one lead to another
assisted by phase-sensitive quartets.

Two additional classes of threshold resonances have been
unveiled by Houzet and Samuelsson \cite{Houzet-Samuelsson} at the
level of the quasiparticle conductance of incoherent MARs. Those threshold resonances are interpreted in terms of higher-order MAR channels involving all three terminals, which open upon reducing bias
voltages. In this case, the ratio $V_a/V_b$ between the bias voltages
can be commensurate or incommensurate.

Already at equilibrium, multiterminal superconducting structures possess striking properties, like Andreev states robustly crossing the zero-energy level \cite{Delft,epjb,Padurariu}.  The potential of such structures is exemplified in an intriguing proposal by Riwar {\it et al.}  \cite{Riwar}: 
it amounts to producing nontrivial topological effects due to Weyl fermions in a four-terminal
all-superconducting structure probed with two small incommensurate voltages.

In contrast with all these predictions, very few experiments are presently available. In a pioneering transport experiment\cite{Francois} on a three-terminal long diffusive junction, clear Josephson-like anomalies have been observed
 at voltages corresponding to quartets emitted by one terminal towards the two others. More recent Shapiro step experiments on the same set-up\cite{Duvauchelle} confirm the coherent character of these resonances. In such an experiment, the currents (and the conductance matrix) are obtained by fixing the voltages $V_a,V_b$. Other experiments are expected in clean systems such as carbon nanotubes or nanowires\cite{Heiblum} defining quantum dots. Moreover, besides experiments controlling both voltages $V_a,V_b$, one should envision experiments setting $V_a=-V_b=V$, which results in $\varphi_Q=\varphi_a+\varphi_b-2\varphi_0$ being a constant of motion. Because of the presence of three terminals, the voltage ($V$) and the phase ($\varphi_Q$) can be taken as two independant variables, and experiments are highly desirable in order to test the dependence of the quartet and MAR current with $V$ and $\varphi_Q$. 

Yet, modeling such multichannel systems poses a formidable
difficulty. The related work by Cuevas and
Pothier\cite{Cuevas-Pothier} is numerically exact, but one contact
being a tunnel one. However, how about a situation in which all
contacts have similar and intermediate transparency, as in the recent
experiment by Pfeffer {\it et al.}\cite{Francois}?  This question
calls for treatments having the capability of dealing with this regime
of intermediate transparency, which requires a fully nonperturbative
approach. An alternative is to focus on a short disordered junction
and use quantum circuit theory\cite{Padurariu,CircuitTheory}, which is
nonperturbative. Yet, this method also raises considerable
difficulties for incommensurate voltages.  More generally, it is
difficult to obtain {\it numerically exact} results when two
incommensurate frequencies and good extended interfaces are involved
in the context of superconducting junctions, and this has not been
done up to now. The paper by Cuevas {\it et al.}\cite{Cuevas-Shapiro}
on Shapiro steps in a one-channel superconducting weak link is one of
the rare examples addressing a problem with two independent
frequencies in the context of nanoscale superconducting junctions.

It is a general trend in the field of quantum electronics that
experiments are controlled by robust effects that, up to now, have
never required supercomputers for being uncovered theoretically at the
semi-quantitative level. It is however true that the challenge of the
exact solution may provide surprises in connection with nonlinear
physics, but experimental relevance of those effects in the context of
multiterminal Josephson junctions has not been proven at present
time. Here, the line of thought is to ignore the effects appearing
solely in the exact solution ({\it e.g.}  not captured by the
approximation used here), which does not prevent us from exploring the exact solution in the future, on the basis of large-scale
numerical calculations.

The present paper proposes a kind of ``unified'' but approximate
numerical framework for multiterminal Josephson junctions, which can
be used for any ratio $V_a/V_b$ between the voltages ({\it e.g.}
commensurate or incommensurate), as a substitute to the unavailable
exact solution. The calculations are based on the so-called
Hamiltonian approach [which was developed by Cuevas,
  Mart\'{\i}n-Rodero and Levy Yeyati\cite{Cuevas} for multiple Andreev
  reflections in a single-channel superconducting weak link], and
intermediate transparencies are used.

\begin{figure*}[htb]
\includegraphics[width=\textwidth]{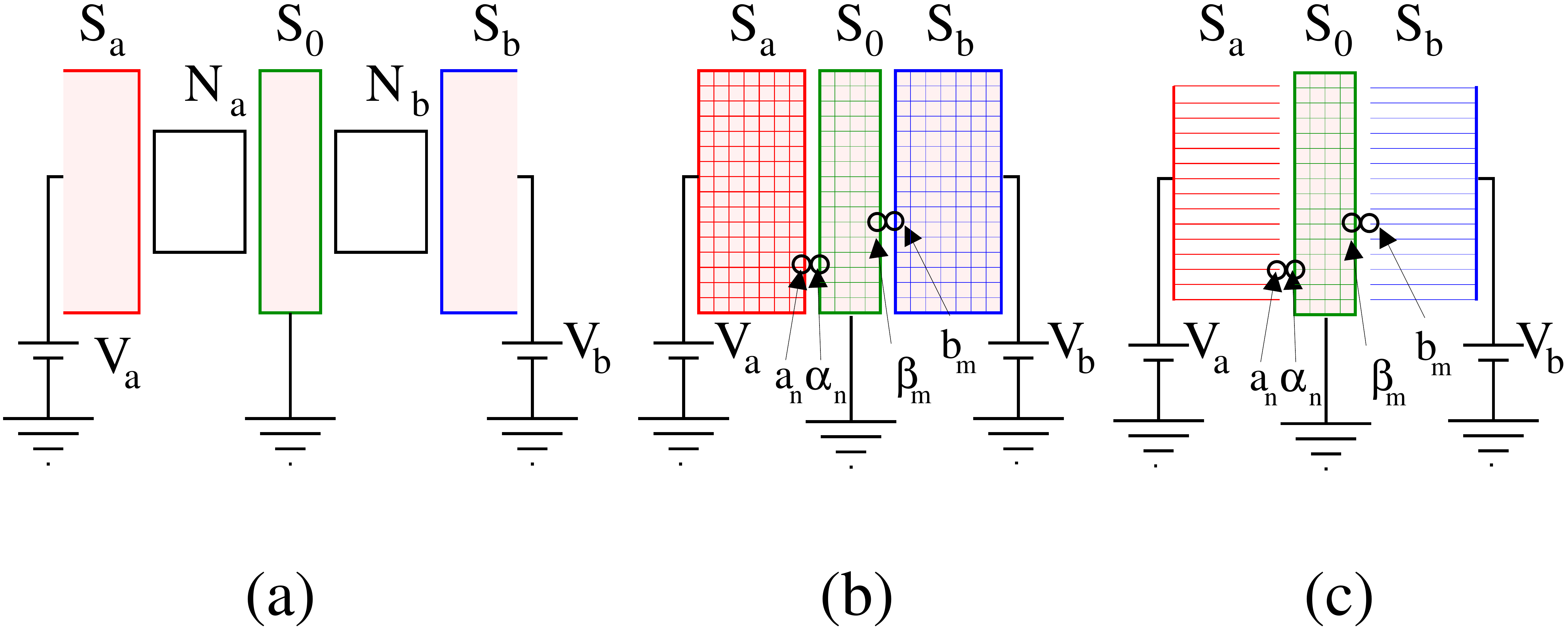}

\caption{Schematics of the three-terminal superconducting
  junction. (a) shows thee same $SNSNS$ set-up as
  Fig.~\ref{fig:artist-view}. (b) shows a two-dimensional
  tight-binding model in which the short normal regions are modeled by
  a tight-binding tunnel amplitude with intermediate transparency. (c)
  shows the considered model in which $S_{a,b}$ are made of a
  collection of one-dimensional channels, thus avoiding the discussion
  of nonlocal effects in the leads $S_{a,b}$.
\label{fig:figure-schema-jonction}
}
\end{figure*}

The method is developed for a three-terminal Josephson junction with
two interfaces having the same intermediate transparency (see
Fig.~\ref{fig:figure-schema-jonction}). It takes into account all
multiple tunneling processes occurring separately at the interfaces
but treats the scattering between interfaces (non-local processes) at
lowest order, as if the distance between the interfaces was larger
than the coherence length at all energies. Notice that this is never
strictly true close to the gap edges where the penetration length of
virtual quasiparticles becomes infinite.

Within this approximation, all resonances at rational $V_a/V_b$ can be
captured in a semi-quantitative way, and the MAR threshold
resonances\cite{Houzet-Samuelsson} are also obtained for irrational
$V_a/V_b$. The method can eventually be generalized to four-terminal
structures.  Agreement with respect to results established previously
\cite{Houzet-Samuelsson,Freyn,Jonckheere} is obtained, and the
numerical results are thoroughly discussed on a physical basis,
especially in connection with the recent experiment by Pfeffer {\it et
  al.}\cite{Francois}. Other points of comparison with known behavior
will be obtained. For instance, in agreement with Jonckheere {\it et
  al.}\cite{Jonckheere}, the current-phase relation has the correct
sign for the first phase-sensitive resonances. In agreement with the
predominance of normal electron transmission over nonlocal Andreev
reflection in a normal metal-superconductor-normal metal junction
($NSN$)\cite{Melin-Feinberg}, the critical current is larger in
absolute value for identical voltages (direct transfer of pairs from
$S_a$ to $S_b$ through $S_c$, coined as\cite{Freyn} ``pair
cotunneling'') than for opposite voltages (the so-called quartets,
also called as double nonlocal Andreev reflection\cite{Freyn}). In
addition, deviations from harmonic current-phase relation are stronger
for larger interface transparency and higher bias voltage. This
reminds of the results of a Blonder-Tinkham-Klapwijk\cite{BTK}
calculation for a $NS$ junction, in which the
subgap current increases with voltage at intermediate transparency.
An interesting physical result on three-terminal junctions will
conclude our study, regarding the relative magnitude of quartets and
multipairs with respect to the MAR signal at intermediate
transparency. Indeed, the multipair and quartet resonances can have a
large ``critical'' current in comparison to the background MAR
current. Therefore they can have a high visibility compared to the
quasiparticle background and lead to a current signal that can be
large enough for experimental detection in a voltage-biased
measurement, which is compatible with the experiment by Pfeffer {\it
  et al.}\cite{Francois}. This conclusion raises the question of
whether multipair resonances also have high visibility in comparison
to quasiparticles in the four-terminal set-up proposed recently by
Riwar {\it et al.}\cite{Riwar}. At least in which region of the
$(V_a,V_b)$ plane is the experimental signal dominated by Weyl
fermion-like effects or by multipair resonances ?

The article is organized as follows. Introductory material on the
method is presented in Sec.~\ref{sec:more-introductory}, all
analytical calculations being relegated to Appendices. The numerical
results are presented and interpreted in Sec.~\ref{sec:theresults}, in
connection to previous
results\cite{Freyn,Houzet-Samuelsson,Jonckheere} and to a recent
experiment by Pfeffer {\it et al.}\cite{Francois}. Concluding remarks
are presented in Sec.~\ref{sec:conclusions}.

\section{Introduction to the method}
\label{sec:more-introductory}
\subsection{The set-up}

Set-ups containing three interfaces are considered in some theoretical
approaches, such as the Usadel equations used by Cuevas and Pothier
\cite{Cuevas-Pothier}, circuit theory treated by Padurariu {\it et
  al.}\cite{Padurariu,CircuitTheory}, or the single quantum dot
connected to three superconductors treated by M\'elin {\it et al.}
(work in progress). Other approaches consider two superconductors
$S_a$ and $S_b$ at voltages $V_a$ and $V_b$ connected to the grounded
$S_0$ (see Fig.~\ref{fig:figure-schema-jonction}), thus with only two
interfaces. This includes Freyn et al.\cite{Freyn}, Jonckheere et
al.\cite{Jonckheere}, with the possibility of a direct coupling
between the two interfaces\cite{epjb}. In the Grenoble experiment
\cite{Francois}, one sample has three interfaces, and another
``reference'' sample close to Fig.~\ref{fig:artist-view} contains only
two $SNS$ junctions. For the latter, the separation between interfaces
is larger than the superconducting coherence length, which explains
why quartets are not visible in the set-up having only two
interfaces. However, there is no technical impossibility to achieve
experimentally another sample with only two interfaces separated by a
few superconducting coherence lengths.  With two interfaces, nonlocal
effects at the scale of the coherence length can be neglected in $S_a$
and $S_b$ but of course not in $S_0$. The avaible experimental data
obtained in this Grenoble experiment \cite{Francois} do not strictly
speaking correspond to the same conditions as those corresponding to
the calculations presented below: the experiment is in the diffusive
limit for a long junction, and the geometry is not identical.

In what follows, the interfaces $S_aS_0$ and $S_bS_0$ are extended
(multichannel), and $S_a$ and $S_b$ are thus separately connected to
$S_0$, forming a three-terminal Josephson junction with only two
interfaces\cite{Freyn} (see
Fig.~\ref{fig:figure-schema-jonction}). The normal metallic regions
$N_a$ and $N_b$ are treated in the short junction limit, and are
approximated by a hopping amplitude $\Sigma_0$, taken for simplicity
as uniform over the junctions.  Strictly speaking, $S_a$, $S_b$ and
$S_0$ are modelled as a collection of independent one-dimensional
channels in which the mean-field BCS superconducting gap is supposed
to be uniform. This assumption is reasonable, even for $S_0$ which is
macroscopic except in the contact region. A decrease of the gap in
$S_0$, due to proximity effect could easily be included in the theory,
but a fully self-consistent calculation in the spirit of
Ref.~\onlinecite{Melin-Bergeret-Yeyati} seems to be out of reach at
present time.

The current $I_{a,\alpha}$ connecting two points $a$ (in $S_a$) and
$\alpha$ (in $S_0$) is calculated (see Appendix~\ref{app:GFs} and
Fig.~\ref{fig:figure-schema-jonction}), then integrated on the contact
area. Within a tunnel Hamiltonian, each elementary tunnel event
connects to only one pair of points $(a,\alpha)$ and the total current
scales linearly with the contact area (or the number of
channels). This holds even if multiple tunneling takes place, as in
the present nonperturbative calculation. In what follows, the currents
per channel are plotted. The calculation uses a tight-binding
Hamiltonian [see
  Eqs.~(\ref{eq:Hamiltonian1})-(\ref{eq:Hamiltonian2})]. Such a
two-dimensional model for a $NSN$ junction in strong nonequilibrium
conditions was treated numerically by M\'elin, Bergeret and Levy
Yeyati\cite{Melin-Bergeret-Yeyati} with recursive Green's function. A
similar assumption was made ({\it e. g.}  neglecting nonlocal effects
in the leads), and this simplifying assumption did not produce
noticeable artifacts on the current. As it was the case for nonlocal
transport at a $NSN$ interface
\cite{Melin-Feinberg-EPJB,Melin-Feinberg,Melin-Bergeret-Yeyati}, the
superconducting electrodes $S_a$ and $S_b$ are modelled as a
collection of one-dimensional channels, connected to the
two-dimensional superconductor $S_0$. Our theoretical approach
\cite{Falci,Melin-Feinberg-EPJB,Melin-Feinberg,Melin-Bergeret-Yeyati}
has proven to be successful to capture the relevant experimental
physics\cite{Beckmann,Russo,Cadden} and this is why we continue here
this theoretical description into the field of three-terminal
Josephson junctions.

The calculation is performed by using as control parameters the
voltages $V_a$, $V_b$ and the phases at the origin of time
$\varphi_a,\varphi_b, \varphi_0$. Contrarily to two-terminal
junctions, these phases play an essential role in DC    transport because
on a line $pV_a+qV_b=0$ ($p,q$ integers), they lead to a constant of
motion $p\varphi_a+q\varphi_b-(p+q)\varphi_0$. This becomes in turn a
potential control parameter in an experiment.

The tight-binding hopping amplitude in the bulk of $S_a$, $S_b$ and
$S_0$ is denoted by $W$. The interface transparency is characterized
by the dimensionless transmission coefficient (in between $0$ and $1$)
and is equal to $T_N\simeq 4\Sigma_0^2/W$ if $\Sigma_0$ is small
compared to $W$ (small interface transparencies). The tunnel junctions
realized in experiments correspond typically to $T_N\simeq
10^{-4}$. Much larger intermediate values $T_N\simeq 0.04\div 0.33$
are used in the following numerical calculations, corresponding to
intermediate interface transparencies. These achieve a compromise
between the presence of a MAR structure that will turn out to develop
[allowing to access a significant number of MAR threshold resonances
  \cite{Houzet-Samuelsson}], and a sufficiently weak subgap
quasiparticle current [allowing multipair resonances in the current to
  have sufficient visibility compared to the quasiparticle background
  current signal].  Notice that the recent experiment by Pfeffer {\it
  et al.}  \cite{Francois} is done with intermediate interface
transparencies (about $0.3$).

\subsection{Symmetries of the current from microscopic calculations}
\label{sec:symetries}
We start by investigating the symmetries of the current, in
order to demonstrate compatibility of the Hamiltonian approach with
expectations based on general symmetry arguments
\cite{Jonckheere}. The corresponding calculations are relegated in the
Appendices. [An introduction to the Green's functions calculations is
  provided in the Appendix~\ref{app:GFs}; an analysis of the symmetries of
  the current is provided in Appendix~\ref{app:sym}]. The strategy is
to obtain symmetries for the bare Green's function, to be
``transferred'' to the fully dressed advanced and retarded Green's
functions, and next to the Keldysh Green's function and to the
current. The conclusion is a confirmation of the arguments by
Jonckheere {\it et al.}\cite{Jonckheere} that the current originates
from two terms with different parities with respect to phase and
voltage reversal. Indeed, it was shown previously\cite{Jonckheere} on the basis of time-reversal
combined to particle-hole symmetry that
$I(V_i,\varphi_j)=-I(-V_i,-\varphi_j)$.  Written in the form of
Eq.~(\ref{eq:2termes-1}), here the current is explicitly decomposed into a term that
is odd in phase -- and thus even in voltage -- (the multipair
component, including the quartet current for opposite voltages), and
into a term that is odd in voltage -- and thus even in phase -- (the
phase-MAR current). In the voltage range considered in what follows, only the contribution to the current which is even in voltage and
odd in phase contributes to the phase-sensitive current at a
multipair resonance: the voltages considered in
Sec.~\ref{sec:theresults} are mostly below threshold for the
appearance of the phase-MAR processes corresponding to the other
parity \cite{Jonckheere}.

\subsection{Perturbative expansion in the number of nonlocal
bare Green's functions}
\label{sec:method-perturb}

A breakdown of perturbation theory in interface transparency for the
current already appears at the level of the
Blonder-Tinkham-Klapwijk\cite{BTK} (BTK) wavefunction calculation of
the Andreev reflection current in a normal
metal-insulator-superconductor ($NIS$) junction. It is underlined that
the following argument is for a single $NIS$ interface, with a
time-independent Hamiltonian. A one-dimensional model of $NIS$
junction is considered now, with a dimensionless interfacial
scattering potential $Z$, according to the standard BTK
notation\cite{BTK}. Perfectly transparent junctions correspond to
$Z=0$, and tunnel contacts to $Z\gg 1$.  An expansion in $1/Z\ll 1$ of
the transmission coefficient is carried out now (namely, a
small-transparency expansion) and, in a second step, the integral over
energy of each term of this expansion will be evaluated from $0$ to
$eV$. This procedure is problematic if $eV$ becomes larger than the
superconducting gap $\Delta$, in which case the gap edge singularity
is within the interval of integration over energy $E$. [In the
  discussion of the BTK calculation, the energy is denoted by $E$ as
  in usual BTK calculations. The symbol $E$ has the same meaning as
  $\omega$ used in the Green's function calculations.]

The BTK amplitudes \cite{BTK} $A(E)$ (Andreev reflection) and $B(E)$
(normal reflection) are expanded as follows in $1/Z$, with $Z\gg 1$:
\begin{eqnarray}
A(E)&=&\frac{u_0^2v_0^2}{(u_0^2-v_0^2)^2Z^4}+
\frac{2u_0^2v_0^2}{(-u_0^2+v_0^2)^3Z^6}\\&+& {\cal
  O}\left(\frac{1}{Z^8}\right)\nonumber\\ 
B(E)&=&1+\frac{1}{Z^2}\left(1-\frac{2u_0^2}{(u_0^2-v_0^2)^2}\right)\\
\nonumber
&+&\frac{1}{Z^4}\left(\frac{3u_0^4}{u_0^2-v_0^2} -
\frac{2u_0^2}{u_0^2-v_0^2}\right)
+{\cal O}\left(\frac{1}{Z^6}\right)
,
\end{eqnarray}
where $u_0$ and $v_0$ denote the BCS coherence factors for electrons
and holes.  Specializing to $E\simeq \Delta$ leads to
\begin{eqnarray}
A(E)&\simeq&
\frac{\Delta^2}{4(E^2-\Delta^2)}\frac{1}{Z^4}\\
\nonumber
&+&\frac{1}{4} \left(\frac{\Delta^2}{E^2-\Delta^2}\right)^{3/2}\frac{1}{Z^6}
+ ...\\
B(E) &\simeq& 1 -\frac{1}{Z^2}
\left(\frac{\Delta^2}{E^2-\Delta^2}\right)^{1/2}\\
\nonumber
&+&\frac{3}{4Z^4} \frac{\Delta^2}{E^2-\Delta^2} + ...
.
\end{eqnarray}
Perturbation theory in $1/Z$ for the spectral current $1+A(E)-B(E)$ is
ill-defined in the window $E\in [\Delta( 1 -\alpha Z^{-4}) , \Delta
  (1+\alpha Z^{-4})]$, where $\alpha$ is a constant of order unity.
Divergences are produced in the energy dependence of the spectral
current, leading to an expression of the current integrated over
energy as the sum of an infinite number of infinite terms, which,
eventually, takes the finite value obtained by BTK \cite{BTK} in
the nonperturbative theory. Thus, perturbative expansions of the
current produce non physical divergences if the gap edge singularities
are included in the energy integration interval.

A similar artifact is present in the forthcoming partially resummed
perturbation theory for a three-terminal junction. In this case, the
exact nonperturbative values of the currents are unreachable, because
of far too long times for their computation. It is emphasized that,
because of multichannel effect, even at a multipair resonance (such as
the quartet resonance), where only one frequency
remains in the calculation, it is not an easy task to obtain the fully
nonperturbative value of the current. On the other hand, the lowest order term
turns out to diverge only logarithmically. This lowest-order term will
be shown to behave correctly with respect to physical expectations and
to the results established over the last few years for the quartet,
multipair\cite{Freyn,Jonckheere} and MAR currents
\cite{Houzet-Samuelsson}. The message is that, in practice, partially
resummed perturbation theory can be used at lowest order to understand
the features of experiments.

The approach relies on the same principle as a previous work by
M\'elin and Feinberg\cite{Melin-Feinberg} in the context of nonlocal
transport in a $NSN$ junction, in connection with a crossover of the
nonlocal conductance from negative to positive as interface
transparency is increased. The idea of the method is to solve exactly
the structure with two interfaces sufficiently remote so that they can
be considered as independent. Then, the current is evaluated as the
interfaces are made closer, but still at a distance of the order of a
few coherence lengths. This amounts to make an expansion in the
strength of the processes connecting both interfaces, which,
technically, relies on an expansion of the current in the number of
back-and-forth amplitudes ({\it e.g.} the nonlocal bare Green's
functions) connecting the interfaces. The technical implementation is
more complex in the case of an all-superconducting device, because of
the time-dependence of the Hamiltonian, but the principle of the
calculation is exactly the same. Such perturbation theory in the
number of nonlocal Green's functions converges well in the work
mentioned above \cite{Melin-Feinberg} on nonlocal transport in a $NSN$
structure: for applied voltage much smaller than the gap,
$\exp[-R/\xi(0)]$ is the small parameter in this expansion even at
high transparency, where $R$ is the separation between interfaces, and
$\xi(0)$ is the zero-energy coherence length. One of the interests of
what follows (going against what may be pessimistically anticipated at
first glance) is to show that physically useful and relevant (even
though not mathematically exact) information can be extracted from a
similar expansion in an all-superconducting set-up. Intermediate
transparency will be used, which has the effect of cutting off
high-order terms in the tunnel amplitude, except in a spectral window
close to the gaps. Roughly speaking, in the all-superconducting case,
the parameter of the perturbative expansion becomes
$\exp[-R/\xi(\omega)]$, which is of order unity if the energy $\omega$
(with respect to the chemical potential of $S_0$) is close to the gap
in the energy integral of the phase-sensitive spectral current: the
finite-energy coherence length $\xi(\omega)$ diverges as
$\omega\rightarrow \Delta$.  Divergences in the superconductor density
of states are also present in the gap edge spectral window, which
explains why the formal breakdown of perturbation in the number of
nonlocal Green's function deserves special discussion for the
considered all-superconducting structure.  Let us underline that this
approach is not equivalent to the tunnel approximation for the $S_0$
interface, as used by Cuevas and Pothier
\cite{Cuevas-Pothier}. Indeed, our calculation treats part of the
multiple scattering events at this interface.  On the contrary, in
Ref.\onlinecite{Cuevas-Pothier}, the superconductors $S_{a},S_{b}$ are
directly coupled in a nonperturbative way through the normal region.

In a more technical language, each microscopic process contributing to
the DC   current at order-$2n$ corresponds to a closed diagram having $2n$
nonlocal bare Green's functions crossing the central superconductor
$S_0$. With this convention, the lowest order in partially resummed
perturbation theory is order-two in the number of nonlocal Green's
functions. The next-order terms is order-four,~...  The Dyson
equations provide a perturbation series in the hopping amplitude
$\Sigma_0$. This series in power of $\Sigma_0$ will be reordered as a
series in the number of nonlocal bare Green's functions, taking into
account to all orders dressing by ``local'' processes [{\it e.g.}
  each of those processes involves only one of the two $S_0S_a$ or
  $S_0S_b$ interfaces, without coupling to the other]. Said
differently in a more physical picture, at lowest order, an
electron-like quasiparticle coming from $S_a$ couples to
MARs to all orders at the $S_aIS_0$ interface. Next, it can experience
electron-hole conversion while crossing $S_0$ only once, and finally,
the resulting hole-like quasiparticle is subject to MARs at the
$S_0IS_b$ interface before being transmitted into $S_b$. Combining
with a similar processes from $S_b$ to $S_a$ leads to physical picture
for the history of the $n=1$ processes contributing to the lowest
order-two approximation for the phase-sensitive current at a multipair
resonance.

\begin{figure*}[htb]
\includegraphics[width=\textwidth]{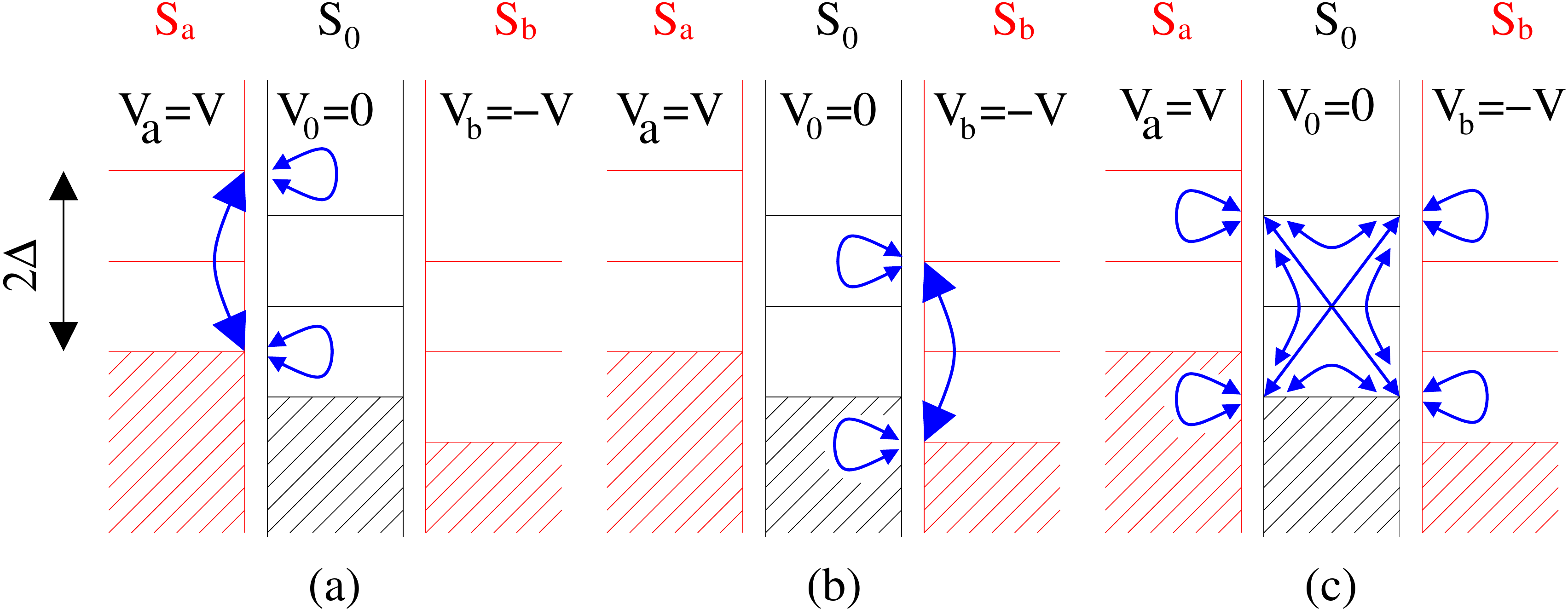}

\caption{Schematics of the three resonances in a three-terminal
  $S_aIS_0IS_b$ junction biased at opposite voltages $V_a=-V_b\equiv
  V$. Resonances (a) and (b) are summed to all orders by the Dyson
  equations. Resonance (c) is not resummed, and it is responsible for
  divergences in perturbation in the number of nonlocal bare Green's
  functions. The arrows denote bare Green's functions. All arrows
  starting and ending at the same energy correspond to
  energy-conserving normal electron transmission, and the arrows
  connecting opposite energies (with respect to the chemical
  potential) correspond to an electron-hole amplitude.
\label{fig:schema-resonances}
}
\end{figure*}

Fig.~\ref{fig:schema-resonances} shows the three processes resonating
at the gap edges, in the simplifying case of opposite bias voltages
(quartet resonance). Both processes on panels a and b take place
``locally'', and they connect energies $\pm \Delta+ eV$ and $\pm
\Delta-eV$ at which the density of states diverges in $S_a$ and $S_b$
respectively. Those divergences need to be regularized by inversion of
the part of the Dyson matrix describing processes taking place locally
at each interface. Conversely, the resonance on
Fig.~\ref{fig:schema-resonances}c will be treated perturbatively, and
it is this resonance which is responsible for divergences in
perturbation in the number of nonlocal Green's functions, if the gap
edge of the superconductor $S_0$ is within the energy integration
interval.  Recipes were attempted for approximate resummations of the
process on Fig.~\ref{fig:schema-resonances}c. Those resummations are
not presented here because it was not possible to estimate the gain in
accuracy on the value of the currents with respect to what
follows. The remainder of the analytical calculations turns out to be
too technical for being presented in the main body of the
article. Appendix~\ref{app:technique-analytique} summarizes those
calculations. Details on the numerical implementation are provided in
Appendix~\ref{app:details-numerics}.

\section{Results}
\label{sec:theresults}

\begin{figure*}[htb]
\includegraphics[width=.8\textwidth]{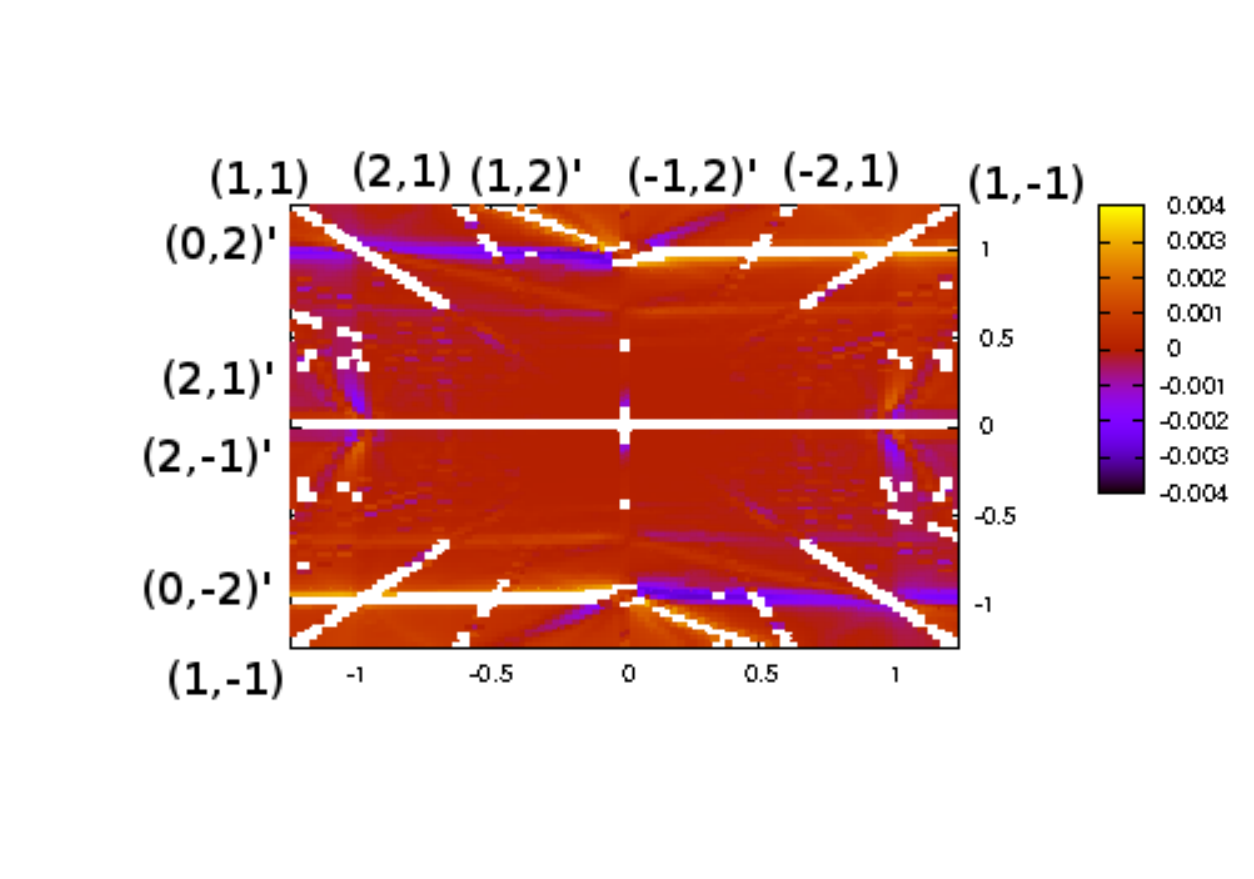}
\caption{Threshold resonances in the nonlocal conductance per channel
  $\partial I_a^{qp}/\partial V_b$, in the $(V_a/\Delta,V_b/\Delta)$
  plane (in units of $e^2/h$ to provide typical orders of magnitude
  with a gap $\Delta=10^{-3}W$, still two orders of magnitude larger
  than a realistic ratio of $\sim 10^{-5}$). The value $N_{max}=6$ is
  used for the cut-off on the number of harmonics of the Josephson
  frequency, and almost indistinguishable data were obtained for
  $N_{max}=7,8,9$. The dimensionless normal state transmission at each
  of the $S_aS_0$ and $S_bS_0$ interfaces is $T_N\simeq 0.33$. The
  separation between the contacts is $R/\xi_0=2$, where $\xi_0$ is the
  low-energy coherence length. The phases are
  $\varphi_a-\varphi_0=\varphi_b-\varphi_0=0$. The white pixels are
  pronounced MAR threshold resonances that fall out of the
  color-scale. The resonant thresholds denoted by $(p,q)$ correspond
  to $p V_a+q V_b=0$ in Eq.~(\ref{eq:reso-HS}), and those denoted by
  $(p',q')'$ correspond to $p' V_a+q' V_b=2\Delta/e$. The resonant
  thresholds $(0,3)'$, $(0,-3)'$, $(1,3)'$ and $(1,-3)'$ are also
  visible on the figure.
\label{fig:colormap-HS}
}
\end{figure*}

As a first point of comparison for partially resummed perturbation
theory, the set of MAR threshold resonances discovered by Houzet and
Samuelsson\cite{Houzet-Samuelsson} was recovered for incommensurate
voltages (however, in the phase-coherent case as far as the present
calculations are concerned). Those threshold resonances correspond to
two families associated to the opening of the channels of MARs:
\begin{equation}
\label{eq:reso-HS}
pV_a+qV_b=0, \mbox{and }
p'V_a+q'V_b=\frac{2\Delta}{e}
,
\end{equation}
with $p,q$ and $p',q'$ four integers. The question arises of whether all
values of ($p,q$) are accessible within lowest order-two in the number
of nonlocal Green's functions. The answer is that a $(p,q)$ diagram
for multipairs can be transformed into a $(p,q)$ diagram for
phase-sensitive MARs by adding one extra line. One easily checks that some multipairs diagrams 
can be constructed by using only two nonlocal lines. Yet, they involve only two split pairs from $S_0$, 
and for higher order processes (sextets, octets..) e.g. if $p>1$ or $q>1$, the other pairs participating to the multipair correlation are not split but originate from direct Andreev reflection at the $N_a$ or $N_b$ interface. 
The above considerations show that at lowest order in the nonlocal processes, all values of
$(p,q)$ are expected to be accounted for, however
approximately. At higher order (not treated hereafter), two, three or higher numbers of Cooper
pairs may split, also accompanied by a $(p,q)$ multipair at both
interfaces. Fig.~\ref{fig:colormap-HS} is a
color-plot showing the dependence on $V_a/\Delta$ and $V_b/\Delta$ of
the nonlocal conductance, the phases being set to zero. This ensures that the multipair current (odd in phase) 
occurring for commensurate voltages is zero and that the calculated features are solely due to MAR quasiparticle transport. With the
intermediate transparency used here ($T_N\simeq 0.33$), those
threshold resonances appear essentially above a voltage range set by
the superconducting gap.

The contact transparencies are also intermediate in the recent
experiment by Pfeffer {\it et al.} on three-terminal metallic
structures \cite{Francois}. The result for the MAR threshold
resonances is thus compatible with the fact that those threshold
resonances were not observed at low bias in this work (due also to inelastic cutoff of high order MAR). 
On the other
hand, this experiment could not probe voltages comparable to the gap,
which would be a requirement for probing the MAR threshold resonances
visible on Fig.~\ref{fig:colormap-HS}.

\begin{figure}[htb]
\centerline{\includegraphics[width=\columnwidth]{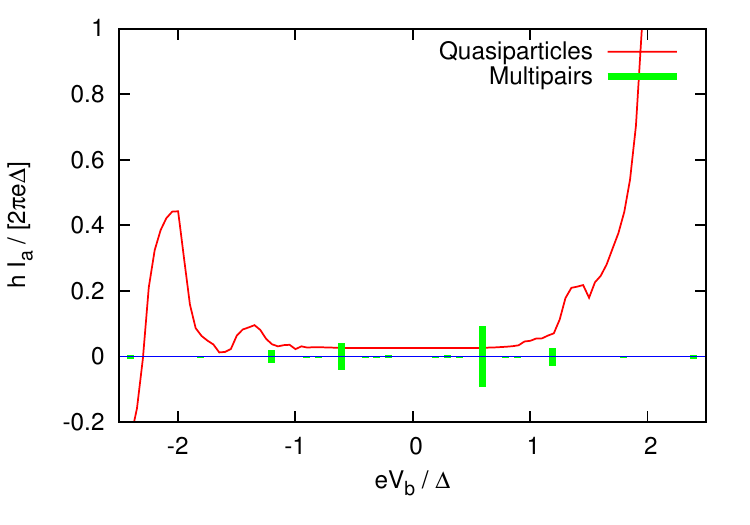}}
\caption{The figure shows the MAR structure in $I_a$ as a function of
  $V_b/\Delta$ for fixed $V_a/\Delta=0.6$, and with
  $N_{max}=6$. Convergence with respect to $N_{max}=7$ was
  verified. The normal state transmission coefficient is $T_N\simeq
  0.33$. All gaps are identical:
  $\Delta\equiv\Delta_a=\Delta_b=\Delta_S=10^{-3}W$. The green lines
  show the value of the critical current of the first multipair
  resonances. The value $R/\xi_0=2$ was used on the figure.
\label{fig:MARS-vs-VB}}
\end{figure}

Interestingly, positive nonlocal conductance $\partial I_a/\partial
V_b$ is obtained in Fig.~\ref{fig:colormap-HS} (see also the
forthcoming Fig.~\ref{fig:MARS-vs-VB}) for some of those MAR threshold
resonances, with the same sign as Cooper pair splitting. This appears
to be reminiscent of the possibility of positive cross-correlations of
MAR currents in a three-terminal all-superconducting hybrid junction
\cite{Duhot}. On the other hand, in a $NSN$ three-terminal junction,
the positive nonlocal conductance at small transparency changes into
negative nonlocal conductance at high transparency
\cite{Melin-Feinberg}, suggesting that, at an oversimplified level
ignoring the condensates while retaining only Cooper pair splitting,
the effective transparency would depend on how remote the voltages are
from threshold resonances. However, this mechanism is not directly
relevant to the calculations presented here because, even at the MAR
threshold resonances, the nonlocal conductance per channel is weak
(see Fig.~\ref{fig:colormap-HS}), thus with underlying microscopic
processes operating in the perturbative regime. The prediction of MAR
threshold resonances with positive nonlocal conductance is interesting
for the prospect of future experiments and theoretical calculations
aiming at an evidence for voltage-controlled positive non
local-conductance.

\begin{figure}[htb]
\centerline{\includegraphics[width=\columnwidth]{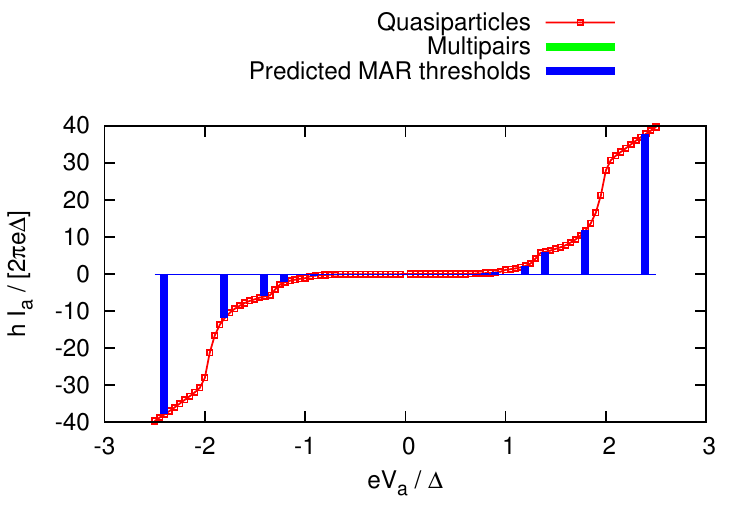}}
\caption{MAR structure in $I_{a,qp}$ as a function of $V_a$ for fixed
  $V_b$, with $N_{max}=6$. Convergence was verified by comparing to
  $N_{max}=7,8,9$. The dimensionless transmission coefficient is
  $T_N\simeq 0.33$. The value $eV_b=0.6\Delta$ is used, as well as
  $\Delta\equiv\Delta_a=\Delta_b=\Delta_0=10^{-3}W$. The multipair
  resonances are also drawn, but they are not visible on this
  scale. The blue impulses show the first MAR threshold
  resonances. The value $R/\xi_0=2$ was used.
\label{fig:MAR1-plein-ecran}}
\end{figure}

\begin{figure}[htb]
\centerline{\includegraphics[width=\columnwidth]{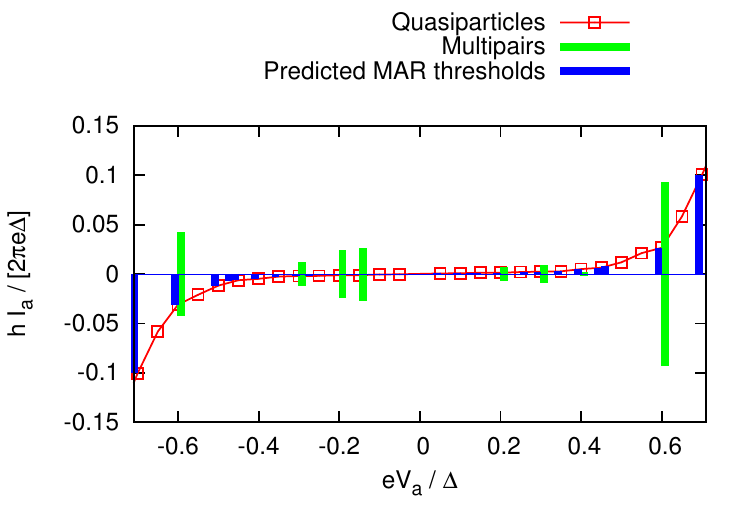}}
\caption{The same as Fig.~\ref{fig:MAR1-plein-ecran}, but now in a
  restricted voltage range. The green impulses show the resonances in
  the phase-sensitive component of the current at $V_a=(p/q)V_b$, with
  $1\le p \le 4$ and $1\le |q| \le 4$. The size of each green impulse
  corresponds to the maximum over the superconducting phase
  $\varphi_0$ of the absolute value of the current at each considered
  multipair resonance. The blue impulses on panel a have the same
  meaning as in Fig.~\ref{fig:MAR1-plein-ecran}, they correspond to
  the MAR threshold resonances \cite{Houzet-Samuelsson}.  The value
  $eV_b=0.6\Delta$ is used, as well as $\Delta\equiv
  \Delta_a=\Delta_b=\Delta_S=10^{-3}W$ and $\Sigma_0/W=0.3$. The value
  $R/\xi_0=2$ was used.
\label{fig:MAR1-zoom}}
\end{figure}

The MAR threshold resonances are better visualized on
Fig.~\ref{fig:MAR1-plein-ecran} featuring the quasiparticle current
$I_{qp}$ as a function of $V_a$ for a fixed $eV_b/\Delta=0.6$, with
$V_a$ and $V_b$ incommensurate. The quasiparticle current as a
function of $V_a$ is dominated by the contribution of the $S_aS_0$
interface, not by the smaller contribution of the processes coupling
the two interfaces. The first MAR threshold resonance
\cite{Houzet-Samuelsson} deduced from the relations (\ref{eq:reso-HS})
coincide with thresholds in the structure for the current plotted as a
function of voltage. The cut-off $N_{max}$ in the space of the
harmonics of the local Green's functions is varied systematically,
keeping only the value $N_{max}=6$ on
Fig.~\ref{fig:MAR1-plein-ecran}. Indistinguishable data were obtained
for $N_{max}=7,8,9$ (not shown here), demonstrating convergence with
$N_{max}$. This excellent convergence is due to the intermediate value
of $T_N=0.33$ between the superconductors.

Fig.~\ref{fig:MAR1-zoom} shows the same data on a more restricted
voltage range smaller than the gap. The multipair resonances are shown
by green impulses on this figure. To obtain those resonances, the
current-phase relation $I_{multipairs}(\varphi_a,\varphi_b)$ was
evaluated as a function of $\varphi_b$ for $\varphi_a=\varphi_0=0$ and
the absolute value of the critical current (maximum value of the
current) was reported on the figure by an impulse. Similarly to a Josephson current, or Shapiro steps, 
a conserved phase variable (here $\varphi_0$ if choosing $\varphi_a=\varphi_b=0$) underlies the multipair current along these green impulses. Multipair
resonances have distinguishing features at $pV_a+qV_b=0$, associated
to the phase-sensitive component of the current ($p=q=1$ corresponds
to the quartets). The value of the multipair critical current at
resonance is much larger than the quasiparticle current, therefore
making the observation of the quartet and multipair resonances
possible at intermediate transparency. The same data are shown on
Fig.~\ref{fig:MARS-vs-VB} as a function of $V_b$ for a fixed value of
$eV_a=0.6 \Delta$. The non-uniform part of the quasiparticle signal on
Fig.~\ref{fig:MARS-vs-VB} corresponds solely to the ``crossed'' or
``nonlocal'' current response of $I_a$ through $S_a$ as a function of
voltage $V_b$ on $S_b$, therefore providing evidence for microscopic
processes coupling nontrivially the two interfaces. More precisely,
the current $I_a$ is calculated according to the approximations
presented in the Appendices, and the derivative in the nonlocal
conductance $G_{a,b}=\partial I_a/\partial V_b$ is evaluated
numerically. The background current (with respect to the multipair
resonances) due to local quasiparticle current at the $S_aS_0$
interface is not small, and local maxima associated to the MAR
threshold resonances are visible. A significant contrast for the
multipair resonances out of the background of MARs is obtained even
for the relatively high value $eV_a/\Delta=0.6$, not being a tiny
fraction of the gap. Those features are consistent with the recent
experiment \cite{Francois} in which resonances were obtained in the
$(V_a,V_b)$ parameter space, consistent with quartets. Higher-order
resonances are not reported in this experiment \cite{Francois},
whereas multipairs have a non-negligible weight in the calculations
presented here. It would be interesting to know whether future
experiments would provide evidence for higher-order resonances.
\begin{figure*}[htb]
\centerline{\includegraphics[width=\columnwidth]{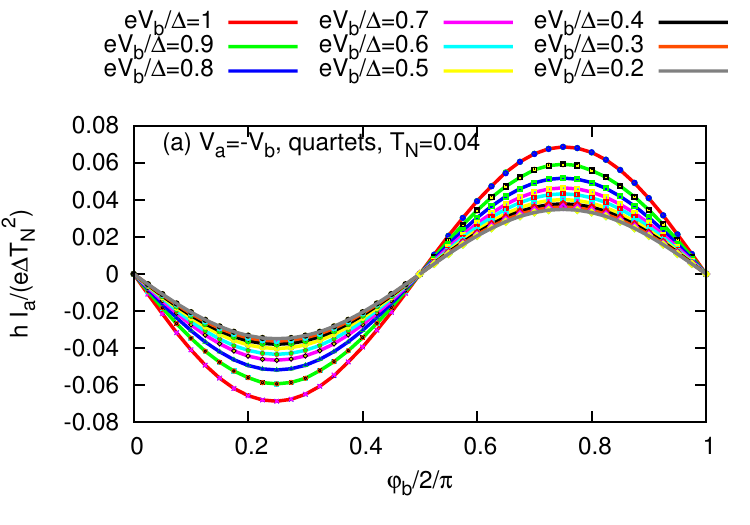}\includegraphics[width=\columnwidth]{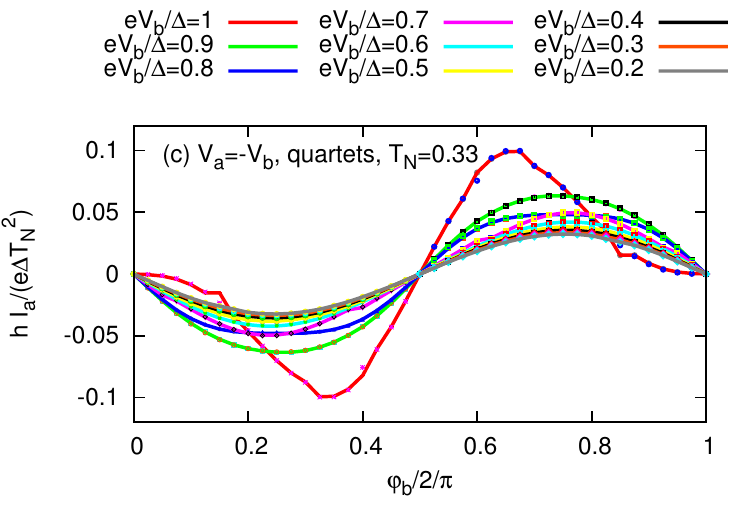}}
\centerline{\includegraphics[width=\columnwidth]{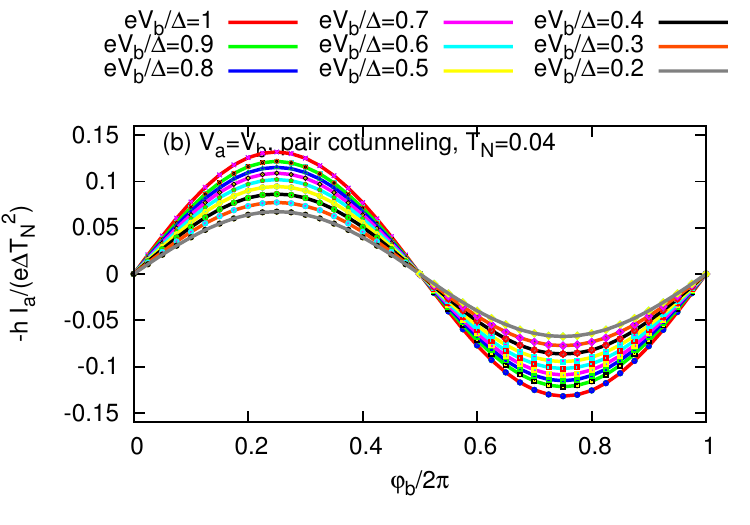}\includegraphics[width=\columnwidth]{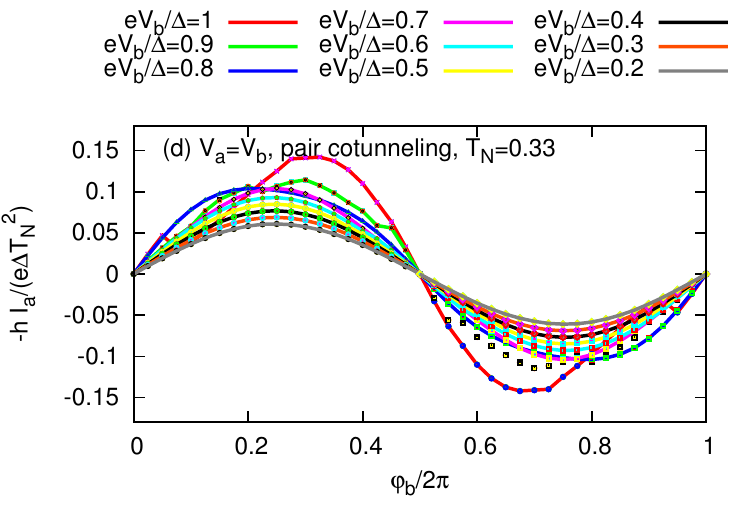}}
\caption{Current-phase relation for $I_a(\varphi_b)$ for $V_a=-V_b$
  (quartets, panels a and c), and $-I_a(\varphi_b)$ for $V_a=V_b$
  (pair cotunneling, panels b and d). The currents are normalized to
  $(e\Delta/h)T_N^2$, which explains that the currents for $T_N\simeq
  0.04$ (panels a and b) are roughly on the same scale as those for
  $T_N\simeq 0.33$ (panels c and d). The figure shows the phase-sensitive
  component of the currents $I_a$ (or $-I_a$) as a function of
  $\varphi_b$ for $\varphi_a=\varphi_0=0$ for the values of $eV_b/\Delta$ shown
  on the figure. All data-points for $N_{max}=4,5,6$ (for $T_N\simeq
  0.04$) and $N_{max}=7,8,9$ (for $T_N\simeq 0.33$) are shown on the
  figure, which provides evidence for excellent convergence as the
  cut-off $N_{max}$ on the number of harmonics of the Josephson
  frequency is increased. The three gaps are identical ($\Delta\equiv
  \Delta_a=\Delta_b=\Delta_S=10^{-3} W$).
\label{fig:Sigma0virgule1}
}
\end{figure*}

Fig.~\ref{fig:Sigma0virgule1} (for $T_N=0.04$ and $T_N=0.33$, and for
$V_a=V_b$ and $V_a=-V_b$) shows the current-phase relations
$I_{multipairs}(\varphi_a,\varphi_b)$ as a function of $\varphi_b$ for
$\varphi_a=\varphi_0=0$, at the resonances associated to $V_a=-V_b$
(quartets\cite{Freyn}, panels a and c), and to $V_a=V_b$ (pair
cotunneling\cite{Freyn} amounting to transferring pairs from $S_a$ to
$S_b$ , or vice-versa, panels b and d). The larger value (in absolute
value) of the critical current of pair cotunneling compared to that of
quartets is at odds with the results obtained by Jonckheere {\it et
  al.}\cite{Jonckheere} for quantum dot calculations in the metallic
regime. If $k_F$ is the Fermi wave-vector and $R$ the distance between
the point contacts, a very specific value $k_F R=2\pi m$ (with $m$ and
integer) was used in this previous work \cite{Jonckheere}. According
to the theory of nonlocal
transport\cite{Falci,Melin-Feinberg,Melin-Feinberg-EPJB}, this
assumption of strict zero-dimensionality is not reliable for extended
interfaces, for which all possible values of $k_F R$ are to be taken
into account [see the form of the nonlocal Green's function in
  Eq.~(\ref{eq:gA-albe}) that is definitely different in the cases of
  generic $k_F R$ and specific value $k_F R=2\pi m$]. This effect on
the magnitude of the critical current for opposite or identical
voltages is due solely to higher-order terms in the tunnel couplings
[a perfect symmetry between the critical currents of quartets and
  elastic cotunneling was obtained by Freyn {\it et al.}\cite{Freyn}
  in the tunnel limit $T_N\ll 1$]. It is concluded that the treatment
of higher order terms in the tunnel amplitudes presented here is
sufficient for obtaining physically sound behavior: the predominance
of pair cotunneling over quartets is reminiscent of that of normal
electron transmission over Cooper pair splitting in a three-terminal
$NSN$ junction, a result first obtained in Ref.
\onlinecite{Melin-Feinberg}.

It is clear from Fig.~\ref{fig:Sigma0virgule1} that the plots of
$I_a(\varphi_b)$ with $\varphi_a=\varphi_0=0$ are compatible with a
$\pi$-shift of the quartet current at opposite bias voltages, absent
for pair cotunneling current at identical bias
voltages. Indeed, the current-phase relation in the adiabatic limit
takes the following form for the quartets:
$I_a=-|I^Q|\sin(\varphi_a+\varphi_b-2\varphi_0)=-|I^Q|\sin(\varphi_b)$. The
minus sign ($\pi$-shift) is a signature of the lowest-order quartet diagram.  Pair cotunneling instead leads
to $-I_a = -|I^{dEC}|\sin(\varphi_a - \varphi_b) = |I^{dEC}|
\sin(\varphi_b)$, and there is no $\pi$-shift in this case. Here, $I^Q$ and $I^{dEC}$ generalize the usual critical current found at equilibrium. Again,
partially resummed perturbation theory is compatible with physical
expectations: the original inversion of the sign of the current-phase
relation for the quartets is related to the electron exchange between
two Cooper pairs in the quartet process \cite{Jonckheere}.

One also notices that the nonharmonic behavior in the current-phase
relation is enhanced as the normal-state transmission increases from
$T_N=0.04$ to $T_N=0.33$ (see Fig.~\ref{fig:Sigma0virgule1}), or as
voltage becomes closer to the gap of $S_0$, which is a behavior
expected on physical grounds.  The proposed interpretation is that the
subgap current at intermediate transparency increases as the bias is
increased towards the superconducting gap, making the junction
effectively more transparent, as in a BTK calculation \cite{BTK} for a
$NS$ interface.

\begin{figure*}[htb]
\begin{minipage}{.69\textwidth}
\hspace*{-2cm}\includegraphics[width=1.1\textwidth]{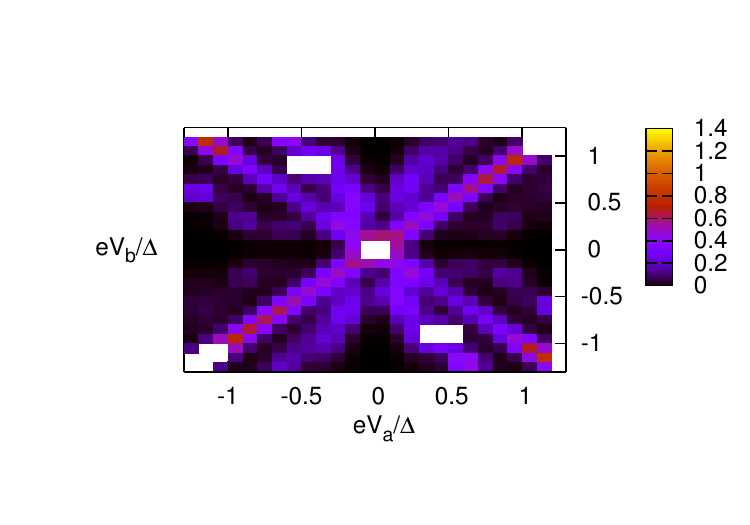}
\end{minipage}
\begin{minipage}{.29\textwidth}
\caption{The figure shows the absolute value of the total quartet critical current
  (in color-scale) as a function of $eV_a/\Delta$ and $eV_b/\Delta$,
  with $N_{max}=9$ as a cut-off on the number of harmonics of the
  Josephson frequency. Indistinguishable figures were obtained for
  $N_{max}=6,7,8$, which demonstrates convergence. The three gaps are
  identical ($\Delta\equiv \Delta_a=\Delta_b=\Delta_S=10^{-3} W$). The
  normal-state transmission coefficient is $T_N=0.33$. The value
  $R/\xi_0=2$ was used on the figure. The current is normalized to
  $(e\Delta/h)T_N^2$. The pixels in white are out of the color-scale.
  \label{fig:colormapN6789}}
\end{minipage}
\end{figure*}

Broadening of the multipair resonances was introduced ``by hand'' in
Fig.~\ref{fig:colormapN6789}, with the (modest) motivation of making
the numerical data look closer to experiment by Pfeffer {\it et al.}
\cite{Francois} where an important broadening is present. The
parameter $\eta$ alone, as it is introduced in our calculations, is
not sufficient for generating such width. Strictly speaking, a finite
width for the resonance in the $(V_a,V_b)$ plane is difficult to
understand in the case of strict voltage bias: as soon as
$V_a+V_b\ne 0$, then $\varphi_a+\varphi_b-2\varphi_0$ starts to
become time-dependent. The finite width is introduced here to mimic
an environment with finite impedance. The color-map in
Fig.~\ref{fig:colormapN6789} underlines similarities and differences
with the experimental data\cite{Francois}. The main difference is that
all superconducting terminals are equivalent in this experiment, and
this is not the case in the set-up on which the present calculations
are carried out. Considering an interpretation of this experiment in
terms of quartets would mean that the latter can be emitted by the
three equivalent superconducting leads in the experiment, according to
the values of the voltages. As a result, three equivalent lines are
obtained experimentally in the $(V_a,V_b)$ plane, which are compatible
with an interpretation in terms of quartets. However, the three
superconducting leads are not equivalent in the set-up used in the
present calculations in which quartets are emitted solely from the
grounded $S_0$, leading to a single resonance line at $V_a+V_b=0$
for the quartets. The other lines in Fig.~\ref{fig:colormapN6789}
correspond to multipair correlations among a larger number of pairs
(sextets, octets,~...)

\section{Conclusions}
\label{sec:conclusions}

An efficient approximate analytical and numerical framework was
provided for a three-terminal Josephson junction. This framework is
based on semi-analytical calculations suitable for providing a kind of
``unified'' picture for the various phenomena taking place in a
three-terminal superconducting junction. The principle of the method
was presented, and it was demonstrated that this method reproduces the
behavior of the various resonances and threshold resonances discovered
over the last few years \cite{Freyn,Houzet-Samuelsson,Jonckheere}, as
well as several features, especially those related to the
current-phase relation. It was demonstrated that MARs and multipairs
can be addressed in the same framework. Calculations with higher
values of interface transparencies are possible in the future with the
same method (at higher computation expenses), suitable for describing
MAR thresholds at higher order.

The proposed method can be an alternative to a frontal numerical
attack to this problem, which seems not to have been attempted up to
now, because of the combined difficulties mentioned above of treating
at once multichannel effects and two independent frequencies. It is
ironic that divergences in tunnel perturbation theory were debated in
the sixties and early seventies, and it is the Keldysh calculations to
all orders carried out by Caroli, Combescot, Nozi\`eres and
Saint-James \cite{Caroli} that solved this debate. The Keldysh method
is applied here to a problem that is sufficiently complex for not
allowing numerically exact solutions to all orders to be carried out
easily. Not unexpectedly, divergences appear in the partially resummed
series, especially with respect to gap edge singularities. The (only
logarithmically divergent) lowest order term in this expansion turns
out to be sufficient for the purpose of discussing the effects that
might be obtained in experiments. Lowest order was benchmarked by
demonstrating compatibility with the results established over the last
years with other methods \cite{Freyn,Houzet-Samuelsson,Jonckheere},
keeping the discussion at the semi-quantitative level. In addition, it
is deduced from the calculations presented above at intermediate
transparency, that the quartet and multipair resonances emerge clearly
from the quasiparticle and multiple Andreev reflection background,
which demonstrates the possibility of experimental observation of
those quartet and multipair resonances.

\begin{figure}[htb]
\centerline{\includegraphics[width=.7\columnwidth]{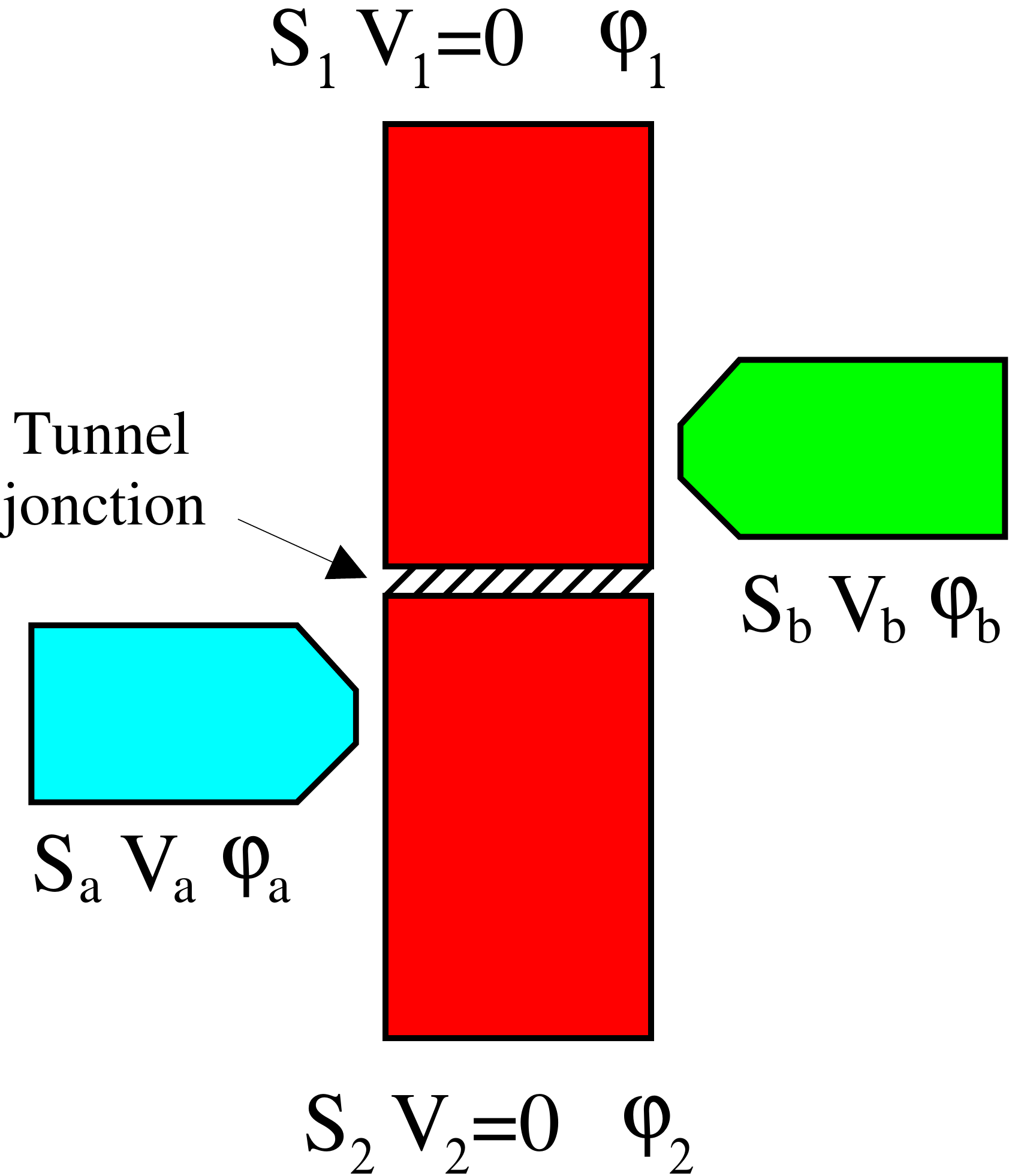}}
\caption{A four-terminal set-up of interest. The separation between
  the $S_1/S_a$ and $S_2/S_b$ contacts is comparable to the coherence
  length.
  \label{fig:device-4T}}
\end{figure}

To conclude, it is suggested now that partially resummed perturbation
theory is scalable to four terminal, and that this planned extension
is promising for addressing a possible interplay between the recently
proposed peculiar features of multiple Andreev reflections at
incommensurate voltages and multipair resonances at commensurate
voltages for nonideal voltage sources. More precisely, we have started
to consider the device in Fig.~\ref{fig:device-4T}, inspired by the
recent preprint by Riwar {\it et al.}\cite{Riwar}. This set-up will be
treated at the order of two nonlocal Green's functions between $S_a$
and $S_b$, crossing the Josephson junction $S_1IS_2$. Interestingly,
the transparency of this Josephson junction is a small parameter for
perturbation theory in the number of nonlocal Green's functions from
$S_a$ to $S_b$, which makes lowest order of the partially resummed
perturbation theory become an exact answer for the four-terminal
set-up shown in Fig.~\ref{fig:device-4T}. The complexity of the code
required to obtain numerically exact results for the four-terminal
structure is the same as for obtaining approximate results in the
three-terminal structure considered here.

\section*{Acknowledgements}
We acknowledge financial support from the French ``Agence National de
la Recherche'' under contract ``Nanoquartets''
12-BS-10-007-04. R.M. used the computer facilities of Institut N\'eel
to develop and run the numerical calculations presented here. It is a
pleasure to thank our colleagues and friends participating to our ANR
project. We wish to express our gratitude to T. Jonckheere, J. Rech
and Th. Martin in Marseille for their collaboration on this
subject. We also wish to thank H. Courtois and F. Lefloch for many
useful discussions related to the quartet state and to their recent
experiment. We also wish to acknowledge financial support and useful
discussions in the framework of Yu. Nazarov's ``Chair of Excellence'',
funded by the ``Fondation Nanoscience'' in Grenoble. We acknowledge in
this context useful discussions on related issues with our colleagues
from CEA-Grenoble: M. Houzet J. Meyer, R. Riwar, X. Waintal. R.M. also
acknowledges a useful discussion with B. Nikolic. Finally, the authors
also thank G\'eraldine Haack for a critical reading and useful
comments on a previous version of our manuscript. 

\appendix

\section{Notations used for Green's functions}
\label{app:GFs}
\subsection{Microscopic Green's functions}
\label{sec:Greens}
The three superconducting electrodes $S_{a_l}$ ($a_l\in\{a,b,0\}$) are
described by the BCS Hamiltonian:
\begin{eqnarray}
\label{eq:Hamiltonian1}
{\cal H}_{a_l}&=&-W \sum_{\langle i,j\rangle} \sum_{\sigma=\uparrow,\downarrow}
\left(c_{i,\sigma}^+ c_{j,\sigma} + h.c.\right))\\
&-&\sum_i \left(\Delta e^{i\varphi_{a_l}}
c_{i,\uparrow}^+ c_{i,\downarrow}^++h.c.\right)
,
\nonumber
\end{eqnarray}
where $\langle i,j\rangle$ is a pair of nearest neighbors on a cubic lattice. The interfaces are coupled by an intermediate hopping amplitude:
\begin{equation}
\label{eq:Hamiltonian2}
{\cal H}_{T,l}=-\Sigma_0 \sum_{\langle a_n,\alpha_n\rangle}\left(c_{a_n,\sigma}^+
c_{\alpha_n,\sigma}+h.c.\right))
.
\end{equation}
The labels $a_n$ and $b_m$ in the Green's functions correspond to the
tight-binding sites at the interfaces of $S_a$ and $S_b$ respectively
(see Fig. \ref{fig:figure-schema-jonction}). The labels $\alpha_n$ and
$\beta_m$ correspond to tight-binding sites in $S_0$, on both
interfaces. The nonlocal bare Green's functions denoted by
$g_{\alpha,\beta}$ connect two generic tight-binding sites $\alpha$
and $\beta$ on the $S_0$ side of both interfaces, separated by a
typical distance $R$ (see Fig.~\ref{fig:figure-schema-jonction}). The
zero-energy Green's function $g_{\alpha,\beta}$ decreases
exponentially with $R$, over the zero-energy coherence length inverse
proportional to the superconducting gap. Two additional labels
associated to precisely which tight-binding site is concerned among
all tight-binding sites present at the interface, have been made
implicit in order to avoid heavy notations. The fully dressed Green's
functions are denoted by $G$. It is supposed in addition that the area
of the multichannel contacts is much smaller than the squared
coherence length.

The choice of the gauge is such that the phases at the origin of time are included in
the superconducting Green's function, and the transitions between
harmonics of the Josephson frequency appear in the tunnel terms at the
interfaces. The local Green's functions take the same form for the
three superconductors:
\begin{eqnarray}
&&\hat{g}_{a_l,a_l}^A(\omega)=\frac{1}{W
\sqrt{(\Delta)^2-(\omega-i\eta)^2}}
\\
&&\left(\begin{array}{cc}-(\omega-i\eta) &
  \Delta\exp(i\varphi_{a_l})\\ \Delta\exp(-i\varphi_{a_l})&
  -(\omega-i\eta) \end{array}\right)
,
\end{eqnarray}
where $a_l\in\{a,b\}$ labels each of the superconductors $S_a$ and
$S_b$, and $a_l\in\{\alpha,\beta\}$ are tight-binding sites belong to
$S_0$, on opposite interfaces. The three superconductors have the same
gap $\Delta$. The phases at $t=0$ in $S_a, S_b$ and $S_0$ are denoted by
$\varphi_a,\varphi_b$ and $\varphi_0$. The parameter $\eta$ can be
viewed as a phenomenological linewidth broadening, introduced as an
imaginary to the energy $\omega$. The parameter $\eta\ll\Delta$ takes
a finite value in the numerical calculations where it plays the role
of regularizing infinities in the partially resummed perturbative
expansions.

The nonlocal Green's function crossing $S_0$ takes the form
\begin{widetext}
\begin{eqnarray}
\label{eq:gA-albe}
\hat{g}_{\alpha,\beta}^A(\omega)=\hat{g}_{\beta,\alpha}^A(\omega)=
\frac{1}{W}\left\{\frac{\cos(k_FR)}{\sqrt{(\Delta)^2-(\omega-i\eta)^2}}
\left(\begin{array}{cc}-(\omega-i\eta) &
  \Delta\exp(i\varphi_0)\\ \Delta\exp(-i\varphi_0)&
  -(\omega-i\eta) \end{array}\right) \right.+ \left. \sin(k_F
R)\left(\begin{array}{cc}1&0\\0&-1\end{array}\right)\right\}.
\end{eqnarray}
\end{widetext}
The retarded Green's functions are obtained from the advanced Green's
functions by changing $\eta$ into $-\eta$.

\subsection{Fourier transform of the Dyson-Keldysh equations}
The Dyson equation in real time for the advanced Green's function is
given by the following convolution:
\begin{eqnarray}
\label{eq:Dyson-time}
\hat{G}^A(t,t')&=&\hat{g}^A(t,t')\\
\nonumber
&+&\int dt_1\hat{g}^A(t,t_1) \hat{\Sigma}(t_1)
\hat{G}^A(t_1,t')
,
\end{eqnarray}
and a similar equation holds for the retarded Green's function. The
notation $\hat{\Sigma}(t)$ in Eq.~(\ref{eq:Dyson-time}) stands for a
diagonal matrix encoding the time-dependent components of the Nambu
hopping amplitudes. Similarly, the Keldysh Green's function is given
by
\begin{eqnarray}
\nonumber
&&\hat{G}^{+,-}(t,t')=\int
dt_1dt_2\left(\hat{I}\delta(t-t_1)+\hat{G}^R(t,t_1) \hat{\Sigma}(t_1)\right)
\\
&\times&\hat{g}^{+,-}(t_1,t_2)\left(\hat{I}\delta(t_2,t') +
\hat{\Sigma}(t_2)\hat{G}^A(t_2,t')\right) .
\label{eq:DK-time}
\end{eqnarray}
Terminal $S_0$ is grounded and terminals $S_a$ and $S_b$ are biased
at voltages $V_a$ and $V_b$, with Josephson frequencies $\omega_a$ and
$\omega_b$ respectively [with $\omega_{a,b}=2eV_{a,b}/\hbar$]. The two
frequencies $\omega_{a,b}$ can be in an arbitrary ratio, commensurate
or incommensurate.

\subsection{Expression of the current}
The current per channel between sites $a$ and $\alpha$ is the sum of
four terms:
\begin{equation}
\label{eq:I-t}
I_{a,\alpha}(t)=\frac{e}{h}\left[{\cal I}_{a,\alpha}^{1,1}(t)
-{\cal I}_{\alpha,a}^{1,1}(t)-{\cal I}_{a,\alpha}^{2,2}(t)
+{\cal I}_{\alpha,a}^{2,2}(t)\right]
,
\end{equation}
with
\begin{eqnarray}
\label{eq:I-t1}
{\cal I}_{a,\alpha}^{1,1}(t)&=& \left[\hat{\Sigma}_{a,\alpha}(t)
  \hat{G}^{+,-}_{\alpha,a}(t,t)\right]^{1,1}\\ \label{eq:I-t2} {\cal
  I}_{\alpha,a}^{1,1}(t)&=& \left[\hat{\Sigma}_{\alpha,a}(t)
  \hat{G}^{+,-}_{a,\alpha}(t,t)\right]^{1,1}\\ \label{eq:I-t3} {\cal
  I}_{a,\alpha}^{2,2}(t)&=& \left[\hat{\Sigma}_{a,\alpha}(t)
  \hat{G}^{+,-}_{\alpha,a}(t,t)\right]^{2,2}\\ \label{eq:I-t4} {\cal
  I}_{\alpha,a}^{2,2}(t)&=& \left[\hat{\Sigma}_{\alpha,a}(t)
  \hat{G}^{+,-}_{a,\alpha}(t,t)\right]^{2,2}
,
\end{eqnarray}
where $G^{+,-}$ is the Keldysh Green's function and
$\hat{\Sigma}_{a,\alpha}(t)$ and $\hat{\Sigma}_{\alpha,a}(t)$ are the
hopping amplitudes for transferring at time $t$ electrons or holes
(according to the selected component of the $2\times 2$ matrix in
Nambu) from $a$ to $\alpha$ or from $\alpha$ to $a$ respectively. The
total current for the extended interface is the sum over all channels:
$I(t)=\sum_n I_{a_n,\alpha_n}(t)$.

\section{Symmetries of the current}
\label{app:sym}
\subsection{Symmetry ${\cal S}_1$: checking that the current
  is a real number}
\label{sec:sym2}
The demonstration is illustrated on an example simpler than the full
Keldysh Green's function. Only one term contributing to the full
$\hat{G}^{+,-}_{\alpha,a}$ is selected as an example:
\begin{eqnarray}
\left[\hat{G}^{+,-}\right]_{\alpha,a}^{example}&=&
\hat{g}^A_{\alpha,\beta}\hat{\Sigma}_{\beta,b}
\hat{g}^A_{b,b} \hat{\Sigma}_{b,\beta} \hat{g}^{+,-}_{\beta,\alpha}
\hat{\Sigma}_{\alpha,a}
\hat{g}^R_{a,a}\\ \left[\hat{G}^{+,-}\right]_{a,\alpha}^{example}
&=&
\hat{g}^A_{a,a} \hat{\Sigma}_{a,\alpha} \hat{g}^{+,-}_{\alpha,\beta}
\hat{\Sigma}_{\beta,b} \hat{g}^R_{b,b} \hat{\Sigma}_{b,\beta}
\hat{g}^R_{\beta,\alpha}
\end{eqnarray}
Then, the following identities are deduced:
\begin{eqnarray}
&&\left[\hat{\Sigma}_{a,\alpha}
    \left[\hat{G}^{+,-}\right]_{\alpha,a}^{example}
    \right]^t\\ \nonumber &&
  =\left[\hat{\Sigma}_{a,\alpha}\hat{g}^A_{\alpha,\beta}\hat{\Sigma}_{\beta,b}
    \hat{g}^A_{b,b} \hat{\Sigma}_{b,\beta}
    \hat{g}^{+,-}_{\beta,\alpha} \hat{\Sigma}_{\alpha,a}
    \hat{g}^R_{a,a}\right]^t\\ \nonumber
  &=&\left(\hat{g}^R_{a,a}\right)^t \hat{\Sigma}_{\alpha,a}
  \left(\hat{g}^{+,-}_{\beta,\alpha}\right)^t \hat{\Sigma}_{b,\beta}
  \left(\hat{g}^A_{b,b}\right)^t \hat{\Sigma}_{\beta,b}
  \left(\hat{g}^A_{\alpha,\beta}\right)^t
  \hat{\Sigma}_{a,\alpha}\\ \nonumber &=&
  \left(\hat{g}^A_{a,a}\right)^* \hat{\Sigma}_{\alpha,a}
  \left(-\hat{g}^{+,-}_{\alpha,\beta}\right)^* \hat{\Sigma}_{b,\beta}
  \left(\hat{g}^R_{b,b}\right)^* \hat{\Sigma}_{\beta,b}
  \left(\hat{g}^R_{\beta,\alpha}\right)^*
  \hat{\Sigma}_{a,\alpha}\\ \nonumber &=&-\left(\hat{g}^A_{a,a}
  \hat{\Sigma}_{a,\alpha} \hat{g}^{+,-}_{\alpha,\beta}
  \hat{\Sigma}_{\beta,b} \hat{g}^R_{b,b} \hat{\Sigma}_{b,\beta}
  \hat{g}^R_{\beta,\alpha} \hat{\Sigma}_{\alpha,a}
  \right)^*\\ \nonumber &=&
  \left[\left[\hat{G}^{+,-}\right]_{a,\alpha}^{example}
    \hat{\Sigma}_{\alpha,a} \right]^* ,
\end{eqnarray}
where the following identity deduced from Eq.~(\ref{eq:gA-albe}) were used:
\begin{equation}
\label{eq:S1}
\mbox{Symmetry ${\cal S}_1$: }
(\hat{g}^A)^t(\omega,\varphi)=(\hat{g}^R)^*(\omega,\varphi) .
\end{equation}
Inserting the frequency variables leads to:
\begin{eqnarray}
&& \left[\hat{\Sigma}_{a,\alpha}^{1,1}(\omega, \omega-\frac{\omega_a}{2})
    \left[\hat{G}^{+,-,11}\right]_{\alpha,a}^{example}
    (\omega-\frac{\omega_a}{2},\omega) \right]^t\\
\nonumber
&&=- \left[
    \left[\hat{G}^{+,-,1,1}\right]_{a,\alpha}^{example} (\omega, \omega-
    \frac{\omega_a}{2})\hat{\Sigma}_{\alpha,a}^{1,1}(\omega-
    \frac{\omega_a}{2},\omega)\right]^* .
.
\end{eqnarray}
It is deduced that
\begin{eqnarray}
&& \left[\hat{\Sigma}_{a,\alpha}^{1,1}(\omega, \omega-\frac{\omega_a}{2})
    \left[\hat{G}^{+,-,11}\right]_{\alpha,a}
    (\omega-\frac{\omega_a}{2},\omega) \right]^t\\
\nonumber
&&=- \left[
    \left[\hat{G}^{+,-,1,1}\right]_{a,\alpha} (\omega, \omega-
    \frac{\omega_a}{2})\hat{\Sigma}_{\alpha,a}^{1,1}(\omega-
    \frac{\omega_a}{2},\omega)\right]^* .
.
\end{eqnarray}
Now, making a change of variable in the integral over $\omega$ leads
to
\begin{eqnarray}
&&   \int d\omega\, \Sigma_a
  \left[\hat{G}^{+,-,t}\right]_{\alpha,a}^{1,1}
  (\omega-\frac{\omega_a}{2},\omega)\\
\nonumber
&=&-\int d\omega\,\Sigma_a
    \left[\hat{G}^{+,-,*}\right]_{a,\alpha}^{1,1}
(\omega+\frac{\omega_a}{2}, \omega)
.
\end{eqnarray}
The notation $\Sigma_a$ stands for the ``11'' component of the Nambu
tunnel amplitude connecting $S_a$ and $S_c$. The components
$I_{\alpha,a}^{1,1}$ and $I_{a,\alpha}^{1,1}$ are given by
\begin{eqnarray}
I_{\alpha,a}^{1,1}&=&\int
d\omega\left[\hat{\Sigma}_{\alpha,a}\hat{G}^{+,-}_{a,\alpha}\right]^{1,1}
\left(\omega,\omega,\varphi_a,\varphi_b\right)\\\nonumber
&=& \int d\omega\,\Sigma_a
\hat{G}^{+,-}_{a,\alpha}
\left(\omega+\frac{\omega_a}{2},\omega,\varphi_a,\varphi_b,\varphi_0\right)\\
I_{a,\alpha}^{1,1}&=& \int
d\omega\left[\hat{\Sigma}_{a,\alpha}\hat{G}^{+,-}_{\alpha,a}\right]^{1,1}
\left(\omega,\omega,\varphi_a,\varphi_b\right)\\
\nonumber
&=&\int d\omega\,\Sigma_a
\hat{G}^{+,-}_{\alpha,a}
\left(\omega-\frac{\omega_a}{2},\omega,\varphi_a,\varphi_b,\varphi_0\right) .
\end{eqnarray}

It is concluded that $(I_{a,\alpha}^{1,1})=-(I_{\alpha,a}^{1,1})^*$,
and $(I_{a,\alpha}^{2,2})=-(I_{\alpha,a}^{2,2})^*$, where the minus
sign arises from the Keldysh Green's function $g^{+,-}$.

\subsection{Symmetry ${\cal S}_2$: changing the sign of the phases}
\label{sec:sym1}
The Green's function given by Eq.~(\ref{eq:gA-albe}) is antisymmetric
under the following transformation:
\begin{equation}
\label{eq:tr}
\mbox{Symmetry ${\cal S}_2$: } g^A(\omega,\varphi)=-\left[\left[g^A(-\omega,
    \pi-\varphi)\right]^*\right]^{11\leftrightarrow 22} .
\end{equation}
The symbol ``$11\leftrightarrow 22$'' means that the ``11'' and ``22''
Nambu components have been exchanged. A similar relation holds for the
Nambu hopping amplitude matrix:
$\hat{\Sigma}=-(\hat{\Sigma}^*)^{11\leftrightarrow 22}$, for the fully
dressed Green's functions: $\hat{G}^A(\omega,\varphi) = -\left[ \left[
    \hat{G}^A (-\omega, \pi - \varphi) \right]^*
  \right]^{11\leftrightarrow 22}$. The demonstration for the Keldysh
Green's function is as follows:
\begin{widetext}
\begin{eqnarray}
&&g^{+,-}(\omega,\varphi)= n_F(\omega)\left[g^A(\omega,\varphi) -
    g^R(\omega,\varphi)\right]\\&=&
  \left[n_F(\omega)-n_F(-\omega)+n_F(-\omega)\right] \left[g^A(\omega,\varphi)
    - g^R(\omega,\varphi)\right]\\ &=&
  -n_F(-\omega)\left[\left(g^A(-\omega,\pi-\varphi)\right)^{*,11
      \leftrightarrow 22} -
    \left(g^R(-\omega,\pi-\varphi)\right)^{*,11\leftrightarrow 22}\right]+
  \left[n_F(\omega)-n_F(-\omega)\right]
  \left[g^A(\omega,\varphi)-g^R(\omega,\varphi)\right]\\ &=&
  -\left[g^{+,-}(-\omega,\pi-\varphi)\right]^{*,11\leftrightarrow 22} +
  \left[n_F(\omega)-n_F(-\omega)\right]
  \left[g^A(\omega,\varphi)-g^R(\omega,\varphi)\right] .
\end{eqnarray}
One obtains the following:
\begin{eqnarray}
&&\int d\omega
  \hat{\Sigma}_{a,\alpha}^{1,1}\left[\hat{G}^{+,-}\right]_{\alpha,a}^{1,1}
  (\omega,\omega,\varphi_a,\varphi_b,\varphi_0)\\ &=&\int
  d\omega\,\hat{\Sigma}_a \left[\left(\hat{I}+\hat{G^R}\hat{\Sigma}\right)
    \hat{g}^{+,-}\left(\hat{I}+\hat{\Sigma}\hat{G}^A\right)
    \right]_{\alpha,a}^{1,1}
  (\omega+\frac{\omega_a}{2},\varphi_a,\varphi_b,\varphi_0)\\ &=& -\int
  d\omega\,\Sigma_a \left\{\left[\left(\hat{I}+\hat{G^R}\hat{\Sigma}\right)
    \hat{g}^{+,-}\left(\hat{I}+\hat{\Sigma}
    \hat{G}^A\right)(-\omega-\frac{\omega_a}{2},-\omega,
    -\varphi_a,-\varphi_b,-\varphi_0)\right]^{*,11\leftrightarrow 22}
  \right\}_{\alpha,a}^{1,1}\\ \nonumber &&+\int d\omega\, \Sigma_a
  \left[\left(\hat{I}+\hat{G^R}\hat{\Sigma}\right)
    \delta\hat{g}^{+,-}\left(\hat{I}+\hat{\Sigma}
    \hat{G}^A\right)\right]_{\alpha,a}^{1,1}(\omega+\frac{\omega_a}{2},\omega
  ,\varphi_a,\varphi_b,\varphi_0)\\ &=&-\int d\omega\,\Sigma_a
  \left\{\left[\left(\hat{I}+\hat{G^R}\hat{\Sigma}\right)
    \hat{g}^{+,-}\left(\hat{I}+\hat{\Sigma}
    \hat{G}^A\right)\right](\omega-\frac{\omega_a}{2},\omega,
  -\varphi_a,-\varphi_b,-\varphi_0) \right\}_{\alpha,a}^{*,2,2}\\
\nonumber &&+\int
  d\omega \,\Sigma_a \left[\left(\hat{I}+\hat{G^R}\hat{\Sigma}\right)
    \delta\hat{g}^{+,-}\left(\hat{I}+\hat{\Sigma}
    \hat{G}^A\right)\right]_{\alpha,a}^{1,1}(\omega+\frac{\omega_a}{2},\omega
  ,\varphi_a,\varphi_b,\varphi_0),
\end{eqnarray}
where
\begin{eqnarray}
\label{eq:AAA1}
g^{+,-}(\omega,\varphi)&=&-\left[g^{+,-}(-\omega,\pi-\varphi)
\right]^{*,11\leftrightarrow 22} + \delta g^{+,-}(\omega,\varphi)\\
\delta g^{+,-}(\omega,\varphi)&=&
\left[n_F(\omega)-n_F(-\omega)\right]
\left[g^A(\omega,\varphi)-g^R(\omega,\varphi)\right]
.
\end{eqnarray}
The following identity was used in order to obtain Eq.~(\ref{eq:AAA1}):
\begin{eqnarray}
g^{+,-}(\omega,\varphi)&=&\left[n_F(\omega)-n_F(-\omega)\right]
\left[g^A(\omega,\varphi)
-g^R(\omega,\varphi)\right]
+ n_F(-\omega)\left[g^A(\omega,\varphi)
-g^R(\omega,\varphi)\right]
\end{eqnarray}
Thus, the total current is given by
\begin{eqnarray}
\nonumber
&&I_{tot}(eV_a,eV_b,\varphi_a,\varphi_b,\varphi_0)= 2\Re
\left[ I_{a,\alpha}^{2,2}(eV_a,eV_b,\varphi_a,\varphi_b,\varphi_0)-
  I_{a,\alpha}^{2,2}(eV_a,eV_b,-\varphi_a,-\varphi_b,-\varphi_0)\right]\\ &+&
2\Re\left[\int d\omega 
  \hat{\Sigma}_{a,\alpha}^{1,1}\left[\left(\hat{I}+\hat{G}^R\hat{\Sigma}\right)
    \delta\hat{g}^{+,-}\left(\hat{I}+\hat{\Sigma}
    \hat{G}^A\right)\right]^{1,1}_{\alpha,a}(\omega+
  \frac{\omega_a}{2},\omega,\varphi_a,\varphi_b,\varphi_0) \right]
.\label{eq:2termes-1}
\end{eqnarray}
\end{widetext}

\section{Sketch of the analytical calculations}
\label{app:technique-analytique}
\subsection{Expansion of the Green's functions in the number of nonlocal bare
Green's functions}
\label{sec:expansion-Dyson}
Now, the approximations are presented. The starting point is the Dyson
equations for the fully dressed Green's functions $\hat{G}$ which take
the form
\begin{eqnarray}
\hat{G}_{\beta,\beta}&=&
\hat{g}_{\beta,\beta}+\hat{G}_{\beta,\alpha}\hat{\Sigma}_{\alpha,a}\hat{g}_{a,a}
\hat{\Sigma}_{a,\alpha}\hat{g}_{\alpha,\beta}\\
\nonumber
&+& \hat{G}_{\beta,\beta}\hat{\Sigma}_{\beta,b}\hat{g}_{b,b} 
\hat{\Sigma}_{b,\beta}\hat{g}_{\beta,\beta}\\
\hat{G}_{\alpha,\beta}&=&
\hat{g}_{\alpha,\beta}+\hat{g}_{\alpha,\alpha}\hat{\Sigma}_{\alpha,a}\hat{g}_{a,a}
\hat{\Sigma}_{a,\alpha}\hat{G}_{\alpha,\beta}\\
\nonumber
&+& \hat{g}_{\alpha,\beta}\hat{\Sigma}_{\beta,b}\hat{g}_{b,b} 
\hat{\Sigma}_{b,\beta}\hat{G}_{\beta,\beta}
.
\end{eqnarray}
Those Dyson equations are next expanded in the number of nonlocal bare
Green's functions. The expression of the Green's functions to order
$(g_{\alpha,\beta})^2$ is the following:
\begin{eqnarray}
\hat{G}_{\alpha,\beta} &=& \hat{A}^{(\alpha,\beta)}_a \hat{g}_{\alpha,\beta} 
\hat{B}^{(\alpha,\beta)}_b + {\cal O}\left(g_{\alpha,\beta}\right)^3\\
\hat{G}_{\beta,\beta}&=&\hat{A}^{(\beta,\beta)}_a
+ \hat{B}^{(\beta,\beta)}_b \hat{g}_{\beta,\alpha} \hat{C}^{(\beta,\beta)}_a
\hat{g}_{\alpha,\beta}\hat{D}^{(\beta,\beta)}_b\\&+& 
{\cal O}\left(g_{\alpha,\beta}\right)^4
\nonumber
,
\end{eqnarray}
with
\begin{eqnarray}
\label{eq:Aaalbe}
\hat{A}^{(\alpha,\beta)}_a&=&\left[\hat{I}- \hat{g}_{\alpha,\alpha}
  \hat{\Sigma}_{\alpha,a} \hat{g}_{a,a}
  \hat{\Sigma}_{a,\alpha}\right]^{-1}\\
\hat{B}^{(\alpha,\beta)}_b&=&
\left[\hat{I} - \hat{\Sigma}_{\beta,b} \hat{g}_{b,b}
  \hat{\Sigma}_{b,\beta} \hat{g}_{\beta,\beta}
  \right]^{-1}\\
\hat{A}^{(\beta,\beta)}_b&=&\left[\hat{I}-
  \hat{g}_{\beta,\beta} \hat{\Sigma}_{\beta,b} \hat{g}_{b,b}
  \hat{\Sigma}_{b,\beta}\right]^{-1} \hat{g}_{\beta,\beta}
\\
\hat{B}^{(\beta,\beta)}_b&=&\left[\hat{I}- \hat{g}_{\beta,\beta}
  \hat{\Sigma}_{\beta,b} \hat{g}_{b,b}
  \hat{\Sigma}_{b,\beta}\right]^{-1}
\\
\hat{C}^{(\beta,\beta)}_a&=&\hat{\Sigma}_{\alpha,a}\hat{g}_{a,a}
\hat{\Sigma}_{a,\alpha} \left[\hat{I}- \hat{g}_{\alpha,\alpha}
  \hat{\Sigma}_{\alpha,a} \hat{g}_{a,a}
  \hat{\Sigma}_{a,\alpha}\right]^{-1}\\
\label{eq:Db}
\hat{D}^{(\beta,\beta)}_b&=&
\left[\hat{I} - \hat{\Sigma}_{\beta,b} \hat{g}_{b,b}
  \hat{\Sigma}_{b,\beta} \hat{g}_{\beta,\beta} \right]^{-1}
,
\end{eqnarray}
where the superscript refers to the overall propagation in the fully
dressed Green's function, and the subscript a or b refers to processes
taking place ``locally'' within each $S_aS_0$ or
$S_bS_0$ interface. Similar expressions are obtained for
$\hat{G}_{\beta,\alpha}$ and $\hat{G}_{\alpha,\alpha}$.

\subsection{Exact expression of the fully dressed Keldysh Green's
  function}
\label{sec:current-Keldysh}
Appendix~\ref{sec:expansion-Dyson} above deals with the expansion of
the advanced and retarded Green's functions. Now, the same expansion
is carried out for the Keldysh Green's function. The first step is to
obtain the specific expression of the Keldysh Green's function for the
three-terminal structure under consideration. The exact fully dressed
Keldysh Green's function is obtained as the sum of 12 terms:
\begin{eqnarray}
&&\hat{\Sigma}_{\alpha,a}\hat{G}^{+,-}_{a,\alpha}
  =\\ &&\hat{F}_1\left[\hat{\Sigma}_{\alpha,a}\hat{g}^{+,-}_{a,a}\nonumber
    \hat{\Sigma}_{a,\alpha}\right]\\\nonumber &+&
  \hat{F}_2\left[\hat{\Sigma}_{\alpha,a}\hat{g}^R_{a,a}\hat{\Sigma}_{a,\alpha}
    \left|\right.
    \hat{\Sigma}_{\alpha,a}\hat{g}^{+,-}_{a,a}\hat{\Sigma}_{a,\alpha}\right]\\ &+&\nonumber
  \hat{F}_3\left[\hat{\Sigma}_{\alpha,a}\hat{g}^R_{a,a}\hat{\Sigma}_{a,\alpha}\left|
    \right.
    \hat{\Sigma}_{\beta,b}\hat{g}^{+,-}_{b,b}\hat{\Sigma}_{b,\beta}\right]\\\nonumber
  &+&\hat{F}_4\left[\hat{\Sigma}_{\alpha,a}\hat{g}^R_{a,a}\hat{\Sigma}_{a,\alpha}
    \hat{g}^{+,-}_{\alpha,\alpha} \right]\\\nonumber
  &+&\hat{F}_5\left[\hat{\Sigma}_{\alpha,a}\hat{g}^R_{a,a}\hat{\Sigma}_{a,\alpha}
    \left|\right.
    \hat{\Sigma}_{\alpha,a}\hat{g}^R_{a,a}\hat{\Sigma}_{a,\alpha}
    \hat{g}^{+,-}_{\alpha,\alpha} \right]\\\nonumber
  &+&\hat{F}_1\left[\hat{\Sigma}_{\alpha,a}\hat{g}^R_{a,a}\hat{\Sigma}_{a,\alpha}
    \hat{g}^{+,-}_{\alpha,\alpha}
    \hat{\Sigma}_{\alpha,a}\hat{g}^A_{a,a}\hat{\Sigma}_{a,\alpha}\right]\\ \nonumber
  &+&\hat{F}_2\left[\hat{\Sigma}_{\alpha,a}\hat{g}^R_{a,a}\hat{\Sigma}_{a,\alpha}
    \left|\right. \hat{\Sigma}_{\alpha,a}\hat{g}^R_{a,a}\hat{\Sigma}_{a,\alpha}
    \hat{g}^{+,-}_{\alpha,\alpha}
    \hat{\Sigma}_{\alpha,a}\hat{g}^A_{a,a}\hat{\Sigma}_{a,\alpha}\right]\\ \nonumber
  &+&
  \hat{F}_3\left[\hat{\Sigma}_{\alpha,a}\hat{g}^R_{a,a}\hat{\Sigma}_{a,\alpha}
    \left|\right. \hat{\Sigma}_{\beta,b}\hat{g}^R_{b,b}\hat{\Sigma}_{b,\beta}
    \hat{g}^{+,-}_{\beta,\beta}\hat{\Sigma}_{\beta,b}\hat{g}^A_{b,b}
    \hat{\Sigma}_{b,\beta}\right]\\ \nonumber &+&
  F_6\left[\hat{\Sigma}_{\alpha,a}\hat{g}^R_{a,a}\hat{\Sigma}_{a,\alpha}\left|\right.
    \hat{\Sigma}_{\beta,b}\hat{g}^A_{b,b}
    \hat{\Sigma}_{b,\beta}\right]\\\nonumber &+&
  \hat{F}_7\left[\hat{\Sigma}_{\alpha,a}\hat{g}^R_{a,a}\hat{\Sigma}_{a,\alpha}
    \left|\right. \hat{\Sigma}_{\alpha,a}\hat{g}^R_{a,a}\hat{\Sigma}_{a,\alpha}
    \left|\right. \hat{\Sigma}_{\beta,b}\hat{g}^A_{b,b}
    \hat{\Sigma}_{b,\beta}\right]\\\nonumber &+&
  \hat{F}_8\left[\hat{\Sigma}_{\alpha,a}\hat{g}^R_{a,a}\hat{\Sigma}_{a,\alpha}
    \left|\right. \hat{\Sigma}_{\beta,b}\hat{g}^R_{b,b}
    \hat{\Sigma}_{b,\beta}\right]\\\nonumber &+& \hat{F}_9
  \left[\hat{\Sigma}_{\alpha,a}\hat{g}^R_{a,a}\hat{\Sigma}_{a,\alpha}
    \left|\right. \hat{\Sigma}_{\beta,b}\hat{g}^R_{b,b}
    \hat{\Sigma}_{b,\beta} \left|\right.
    \hat{\Sigma}_{\alpha,a}\hat{g}^A_{a,a}\hat{\Sigma}_{a,\alpha}\right]
  ,
\end{eqnarray}
with
\begin{eqnarray}
\hat{F}_1\left[\hat{X}\right]&=&\hat{X}
\hat{G}_{\alpha,\alpha}^A\\ \hat{F}_2\left[\hat{X}\left|\right.\hat{Y}\right]&=&\hat{X}
\hat{G}_{\alpha,\alpha}^R\hat{Y}\hat{G}^A_{\alpha,\alpha}\\ \hat{F}_3\left[\hat{X}\left|\right.\hat{Y}\right]&=&\hat{X}
\hat{G}_{\alpha,\beta}^R\hat{Y}\hat{G}^A_{\beta,\alpha}\\ \hat{F}_4\left[\hat{X}\right]&=&\hat{X}\\ \hat{F}_5\left[\hat{X}\left|\right.\hat{Y}\right]&=&\hat{X}
\hat{G}^R_{\alpha,\alpha}\hat{Y}\\ \hat{F}_6\left[\hat{X}\left|\right.\hat{Y}\right]&=&
\hat{X}\hat{g}^{+,-}_{\alpha,\beta}\hat{Y}\hat{G}^A_{\beta,\alpha}\\ \hat{F}_7\left[\hat{X}\left|\right.\hat{Y}\left|\right.\hat{Z}\right]&=&
\hat{X}\hat{G}^R_{\alpha,\alpha}\hat{Y}\hat{g}^{+,-}_{\alpha,\beta}
\hat{Z}\hat{G}^A_{\beta,\alpha}\\ \hat{F}_8\left[\hat{X}\left|\right.\hat{Y}\right]&=&
\hat{X}\hat{G}^R_{\alpha,\beta}\hat{Y}\hat{g}^{+,-}_{\beta,\alpha}\\ \hat{F}_9\left[\hat{X}\left|\right.\hat{Y}\left|\right.\hat{Z}\right]&=&
\hat{X}\hat{G}^R_{\alpha,\beta}\hat{Y}\hat{g}^{+,-}_{\beta,\alpha}
\hat{Z}\hat{G}^A_{\alpha,\alpha} ,
\end{eqnarray}
where the matrices $\hat{X}$, $\hat{Y}$ and $\hat{Z}$ do not couple the two
interfaces.

\subsection{Expansion of the Keldysh Green's function in the number
of nonlocal bare Green's functions}
\label{sec:kFR}
In the next step, all of the fully dressed advanced and retarded
Green's functions in $\hat{F}_n$ ($n=1...9$) are expanded in
perturbation in the number of nonlocal bare Green's function. The
final expression of the Keldysh Green's functions first involves the
matrix elements of some products between the matrices $\hat{X}$,
$\hat{Y}$ and $\hat{Z}$, and the matrices $\hat{A}$, $\hat{B}$,
$\hat{C}$ and $\hat{D}$.  Each of the bare nonlocal bare Green's
functions
\begin{equation}
\label{eq:tr-mode}
\langle\langle \hat{g}_{\alpha,\beta}\left(\alpha^1,N_a^1
\left|\right.\alpha^2, N_b^2\right) 
\hat{g}_{\beta,\alpha}\left(\alpha^3,N_b^3
\left|\right.\alpha^4, N_a^4\right) \rangle\rangle
\end{equation}
is evaluated from Fourier transform over times $t$ and
$t'$:
\begin{eqnarray}
\nonumber &&\hat{g}_{\alpha,\beta}\left(\alpha^1,N_a^1
\left|\right.\alpha^2, N_b^2\right)\\ &=& \int dt dt'
\exp{\left\{i\left[\omega+N_a^1
    \frac{\omega_a}{2}\right]t\right\}}\\ &\times& \nonumber
\exp{\left\{-i\left[\omega+N_b^2 \frac{\omega_b}{2}\right]t'\right\}}
\hat{g}_{\alpha,\beta}(t'-t)\\ &=& \delta_{\left[N_a^1
    \frac{\omega_a}{2}\right]- \left[N_b^2 \frac{\omega_b}{2}\right]}
\hat{g}_{\alpha,\beta}\left(\omega+N_a^1 \frac{\omega_a}{2}\right) ,
\label{eq:truc1}
\end{eqnarray}
where $\alpha^n$ is a Nambu label. The quantities $N_a^1$ and $N_b^2$
encodes multiples of the voltage frequencies $\omega_a/2\equiv e V_a$
and $\omega_b/2\equiv e V_b$ in electrodes $S_0S_a$ and
$S_0S_b$. Eq.~(\ref{eq:truc1}) is valid for arbitrary values of
$\omega_a$ and $\omega_b$, not necessarily commensurate.

\begin{widetext}œ
Coming back to the expansion of the Keldysh function at the lowest
order-two, an example is the following:
\begin{eqnarray}
\hat{F}_2\left[\hat{X}\left.\right|\hat{Y}\right] &=&
\hat{X}\hat{A}^{(\alpha,\alpha),R}\hat{Y}\hat{A}^{(\alpha,\alpha),A} +
\hat{X}\hat{A}^{(\alpha,\alpha),R}\hat{Y}\hat{B}^{(\alpha,\alpha),A}
\hat{g}_{\alpha,\beta}^A\hat{C}^{(\alpha,\alpha),A}
\hat{g}_{\beta,\alpha}^A \hat{D}^{(\alpha,\alpha),A}\\ &+&
\hat{X}\hat{B}^{(\alpha,\alpha),R} \hat{g}_{\alpha,\beta}^R
\hat{C}^{(\alpha,\alpha),R} \hat{g}_{\beta,\alpha}^R
\hat{D}^{(\alpha,\alpha),R}\hat{Y} \hat{A}^{(\alpha,\alpha),A} +{\cal
  O}\left((g_{\alpha,\beta})^4\right) .  \nonumber
\end{eqnarray}
Expanding all matrix products leads to
\begin{eqnarray}
\label{eq:a1}
&&\left\{\hat{F}_2\left[\hat{X}\left.\right|\hat{Y}\right]\right\}_{0,0}
=
\left\{\hat{X}\hat{A}^{(\alpha,\alpha),R}\hat{Y}\hat{A}^{(\alpha,\alpha),A}
\right\}_{0,0}
\label{eq:a2}\\\nonumber
&+& \sum_{\kappa_1,\kappa_2,\kappa_3}
\left\{\hat{X}\hat{A}^{(\alpha,\alpha),R}\hat{Y}\hat{B}^{(\alpha,\alpha),A}
\right\}_{0,\kappa_1} \times
\left\{\hat{C}^{(\alpha,\alpha),A}\right\}_{\kappa_2,\kappa_3}
\left\{\hat{D}^{(\alpha,\alpha),A}\right\}_{\kappa_4,0} \langle\langle
\left\{\hat{g}_{\alpha,\beta}^A \right\}_{\kappa1,\kappa2}
\left\{\hat{g}_{\beta,\alpha}^A\right\}_{\kappa_3,\kappa_4}\rangle\rangle
\label{eq:a3}\\\nonumber
&+&\sum_{\kappa_1,\kappa_2,\kappa_3}
\left\{\hat{X}\hat{B}^{(\alpha,\alpha),R}\right\}_{0,\kappa_1}
\left\{\hat{C}^{(\alpha,\alpha),R}\right\}_{\kappa_2,\kappa_3}
\left\{\hat{D}^{(\alpha,\alpha),R}\hat{Y}
\hat{A}^{(\alpha,\alpha),A}\right\}_{\kappa_4,0} \langle\langle
\left\{ \hat{g}_{\alpha,\beta}^R\right\}_{\kappa_1,\kappa_2} \left\{
\hat{g}_{\beta,\alpha}^R\right\}_{\kappa_3,\kappa_4}
\rangle\rangle\\\nonumber &+&{\cal O}\left((g_{\alpha,\beta})^4\right)
.
\end{eqnarray}
The variables $\kappa_n$ (with $n=1,2,3$) label both Nambu and the
harmonics of the Josephson frequency. The average over the Friedel
oscillations in Eq.~(\ref{eq:tr-mode}) may appear as a standard
procedure within the theory of nonlocal transport in three-terminal
$NSN$ structures \cite{Falci,Melin-Feinberg,Melin-Feinberg-EPJB}:
\begin{equation}
\label{eq:kFR-average}
\langle\langle F\rangle\rangle(R)= \frac{k_F}{2\pi}
\int_{R-\pi/k_F}^{R+\pi/k_F}  F(r) dr
,
\end{equation}
where the function $F(r)$ has rapid Friedel oscillations at the scale
of the Fermi wavelength $\lambda_F=2\pi/k_F$ (to be averaged out
because of multichannel averaging), but its envelope decays smoothly
on the much longer scale of the coherence length $\xi$. The Nambu
elements of the non-local Green's function
$\hat{g}_{\alpha,\beta}^{\epsilon_1}(R)$ can be written as
\begin{eqnarray}
\nonumber &&\hat{g}_{\alpha,\beta}^{\epsilon_1}(R)= \frac{f(R)}{W}
\exp\left\{-\frac{R}{\xi\left(\omega-i\eta \epsilon_1\right)}\right\}
\left\{\frac{\cos{(k_F
    R)}}{\sqrt{\Delta^2-(\omega-i\eta\epsilon_1)^2}}
\left[\begin{array}{cc} -(\omega-i\eta\epsilon_1)&\Delta\\ \Delta &
    -(\omega-i\eta\epsilon_1)\end{array}\right]\right.\\ &&\left.+\sin(k_F
R)\left[\begin{array}{cc} 1 & 0 \\ 0 & -1\end{array}\right] \right\} ,
\end{eqnarray}
\end{widetext}
where $\epsilon_1=\pm$ labels advanced or retarded. The exponent in
its power-law decay is dimension-dependent, and $\xi(\omega)$ is the
energy-dependent coherence length. The parameter $\eta$ is a small
linewidth broadening. The function $f(R)$ is a prefactor that does
not oscillate at the scale of the Fermi wave-length. It will be taken
as a constant in our numerical calculations, because the contacts are
at distance large compared to their size.

Using $\langle\langle \cos^2(k_F R)\rangle\rangle = \langle\langle\sin^2(k_F
R)\rangle\rangle=1/2$ and $\langle\langle \cos(k_FR)\sin(k_F
R)\rangle\rangle=0$ leads to 
\begin{widetext}
\begin{eqnarray}
\label{eq:average}
&&\langle\langle \hat{g}_{\alpha,\beta}^{\epsilon_1}\otimes
\hat{g}_{\beta,\alpha}^{\epsilon_2} \rangle\rangle=
\frac{1}{2}\left(\frac{f(R)}{W}\right)^2
\exp\left\{-\frac{R}{\xi(\omega-i\eta\epsilon_1)}\right\}
\exp\left\{-\frac{R}{\xi(\omega-i\eta\epsilon_2)}\right\}\\ \nonumber
&\times&\left\{ h_{\epsilon_1,\epsilon_2}(\omega_0,\tilde{\omega}_0)
\left[\begin{array}{cc}-(\omega_0-i\eta\epsilon_1)&\Delta\\ \Delta&-(\omega_0-i\eta\epsilon_1)\end{array}\right]_{\rm
  Nambu} \otimes
\left[\begin{array}{cc}-(\tilde{\omega}_0-i\eta\epsilon_2)&\Delta\\ \Delta&-(\tilde{\omega}_0-i\eta\epsilon_2)\end{array}\right]_{\rm
  Nambu} \right.  \\&&+\left.
\left[\begin{array}{cc}1&0\\0&-1\end{array}\right]_{\rm Nambu} \otimes
\left[\begin{array}{cc}1&0\\0&-1\end{array}\right]_{\rm Nambu}
\right\} ,\nonumber
\end{eqnarray}
\end{widetext}
with
\begin{equation}
h_{\epsilon_1,\epsilon_2}(\omega_0,\tilde{\omega}_0)=
\frac{1}{\sqrt{\Delta^2-(\omega_0-i\eta\epsilon_1)^2}
  \sqrt{\Delta^2-(\tilde{\omega}_0-i\eta\epsilon_2)^2}} ,
\end{equation}
with
\begin{eqnarray}
N_a^1\omega_a&=&N_b^2\omega_b\\
N_b^3\omega_b&=&N_a^4\omega_a
\end{eqnarray}
and with $\omega_0=\omega+N_a^1\omega_a/2$ and $\tilde{\omega}_0
=\omega+N_b^3\omega_b$.

\section{Details on the numerics}
\label{app:details-numerics}
The expansion in the number of nonlocal bare Green's functions
sketched in the previous Appendices was combined to an algorithm for
evaluating the local Green's functions dressed by processes taking
place ``locally'' at each $S_aS_0$ or $S_bS_0$ interface. All
components of the Green's functions connecting any harmonics to any
other are required to be evaluated in order to obtain the fully
dressed Keldysh Green's function of the three-terminal $S_aS_0S_b$
structure, approximated as above . However, inversions of matrices connecting both interfaces
are never required to be performed, which explains the reduced
computation times. Direct matrix inversion of matrices defined locally
at a single interface and thus not coupling the two interfaces is
numerically efficient if the cut-off in the number of harmonics of the
Josephson frequency is not larger than $\sim 10$, which is the case in
what follows because the calculations are restricted to voltage
$eV\agt \Delta/10$. Otherwise, if needed, a possible future version of
the code will evaluate the fully dressed local Green's functions from
one-dimensional recursive Green's functions in energy \cite{Cuevas},
allowing for the possibility of addressing voltages smaller than
$eV\alt \Delta/10$.

The size of the matrices to be inverted would be
\begin{equation}
\label{eq:brutal}
[2(2N_{max}+1)^2] \times [2(2N_{max}+1)^2],
\end{equation}
if all information about nonlocal processes would be kept in a frontal
approach. Programming the problem in this way was attempted, but
running codes with increasing values of $N_{max}$ upon reducing
voltage becomes undoable with standard computer resources, even if the
voltages are not necessarily small compared to the gap. In addition,
the prohibitive computation time given by Eq.~(\ref{eq:brutal}) has to
be multiplied by a factor $N_{k_F R}$, in order to calculate the
integral over the Friedel oscillations with microscopic phase factor
$k_F R$ suitable for a multichannel contact (see
Eq.~\ref{eq:kFR-average}). Alternatively, the spatial aspect of this
problem may be treated with recursive Green's functions in real space,
extended to include the harmonics of half the Josephson frequency
(below a given voltage-sensitive cut-off), but this numerical
implementation appears to be even more computationally demanding than
an average over $k_F R$ for a collection of independent channels.  A
scaling factor in the computation time, of typical value
$N_\omega={\cal O}(10^3)$, is also required for the evaluation of the
integral over energy of the spectral current (containing sharp
resonances) with an adaptative algorithm.

The size of the matrices to be inverted with partially resummed
perturbation theory used here is only
\begin{equation}
\label{eq:our-algorithm}
[2(2N_{max}+1)]\times [2(2N_{max}+1)]
,
\end{equation}
not larger than for a single interface. In addition, the average over
the Friedel oscillations is built-in in the case of the expansion in
the number of nonlocal bare Green's functions, and there is thus a
gain of computation time by the factor $N_{k_F R}$, because, at the
lowest order-two in the partially resummed perturbation theory, the
multichannel junction maps onto a single-channel problem. Again,
$N_\omega={\cal O}(10^{3})$ calls to the spectral current routine are
required in order to integrate it over energy with an adaptative
algorithm. Then, the numerical calculations become approximate, but,
above all, they become doable without running them on a
supercomputer.


\begin{thebibliography}{99}
\bibitem{Cuevas-Pothier} J. C. Cuevas and H. Pothier, Phys. Rev. B
  {\bf 75}, 174513 (2007).
\bibitem{Jillie} N. A. H. Nerenberg, J. A. Blackburn, and D. W. Jillie, Phys. Rev. B {\bf 21}, 118 (1980).
\bibitem{Freyn} A. Freyn, B. Doucot, D. Feinberg, and R. M\'elin,
  Phys. Rev. Lett. {\bf 106}, 257005 (2011).
\bibitem{Jonckheere} T. Jonckheere, J. Rech, T. Martin, B. Dou\c{c}ot,
  D. Feinberg, R. M\'elin, Phys. Rev. B {\bf 87}, 214501 (2013).
\bibitem{Houzet-Samuelsson} M. Houzet and P. Samuelsson, Phys. Rev. B
  {\bf 82}, 06051 (2010).
\bibitem{Delft} B. van Heck, S. Mi, and A.R. Akhmerov, Phys. Rev. B
{\bf 90}, 155450 (2014).
\bibitem{epjb} D. Feinberg, T. Jonckheere, J. Rech, T. Martin, B. Dou\c{c}ot, and R. M\'elin, Eur. Phys. J. B {\bf 88}, 99 (2015). 
\bibitem{Padurariu} C. Padurariu, T. Jonckheere, J. Rech, R. Mélin,
  D. Feinberg, T. Martin, Yu. V. Nazarov, Phys. Rev. B {\bf 92}, 205409 (2015).
\bibitem{Riwar} R.-P. Riwar, M. Houzet, J.S. Meyer, Y.V. Nazarov,
  arXiv:1503.06862
\bibitem{Francois} A.H. Pfeffer, J.E. Duvauchelle, H. Courtois,
  R. M\'elin, D. Feinberg, F. Lefloch, Phys. Rev. B {\bf 90}, 075401
  (2014).
\bibitem{Duvauchelle} J.E. Duvauchelle et al., in preparation (2016).
\bibitem{Heiblum} M. Heiblum, private communication.
\bibitem{CircuitTheory} C. Padurariu, T. Jonckheere, J. Rech, T. Martin, R. M\'elin, and D. Feinberg, in preparation (2016).
\bibitem{Cuevas} J. C. Cuevas, A. Mart\'{\i}n-Rodero, and A. Levy Yeyati,
  Phys. Rev.  B{\bf 54}, 7366 (1996).
\bibitem{Cuevas-Shapiro} J.C. Cuevas, J. Heurich,
  A. Mart\'{\i}n-Rodero, A. Levy Yeyati and G. Sch\"on,
  Phys. Rev. Lett. {\bf 88}, 157001 (2002).
\bibitem{Melin-Feinberg} R. M\'elin and D. Feinberg, Phys. Rev. B {\bf
  70}, 174509 (2004).
\bibitem{BTK} G.E. Blonder, M. Tinkham and T.M. Klapwijk, Phys. Rev. B
  {\bf 25}, 4515 (1982).
\bibitem{Melin-Bergeret-Yeyati} R. M\'elin, F.S. Bergeret, A. Levy
  Yeyati, Phys. Rev. B {\bf 79}, 104518 (2009).
\bibitem{Melin-Feinberg-EPJB} R. M\'elin and D. Feinberg,
  Eur. Phys. J. B {\bf 26}, 101 (2002).
\bibitem{Falci} G. Falci, D. Feinberg, and F. W. J. Hekking,
  Europhys. Lett.  {\bf 54}, 255 (2001).
\bibitem{Beckmann} D. Beckmann, H. B. Weber, and H. v. L\"ohneysen,
  Phys. Rev.  Lett. {\bf 93}, 197003 (2004)
\bibitem{Russo} S. Russo, M. Kroug, T. M. Klapwijk, and
  A. F. Morpurgo, Phys.  Rev. Lett. {\bf 95}, 027002 (2005).
\bibitem{Cadden} P. Cadden-Zimansky and V. Chandrasekhar,
  Phys. Rev. Lett. {\bf 97}, 237003 (2006); P. Cadden-Zimansky,
  Z. Ziang, and V. Chandrasekhar, New J. Phys. {\bf 9}, 116 (2007).
\bibitem{Duhot} S. Duhot, F. Lefloch, M. Houzet, Phys. Rev. Lett. {\bf
  102}, 086804 (2009).
\bibitem{Caroli} C.  Caroli, R.  Combescot, P.  Nozi\`eres, and D.
  Saint-James, J. Phys. C {\bf 4}, 916 (1971); {\bf 5}, 21 (1972).
\end{thebibliography}
\end{document}